\documentclass[12pt]{article}

\usepackage{amsmath,amssymb,feynmf}

\def\a{\alpha}
\def\b{\beta}

\def\d{\delta}

\def\f{\phi}

\def\j{\psi}
\def\k{\kappa}

\def\m{\mu}
\def\n{\nu}

\def\q{\theta}
\def\r{\rho}
\def\s{\sigma}

\def\F{\Phi}
\def\G{\Gamma}

\def\Ld{\Lambda}

\def\S{\Sigma}

\allowdisplaybreaks[4]

\renewcommand{\title}[1]{\null\vspace{10mm}\noindent
                         {\Large{\bf #1}}\vspace{10mm}}
\newcommand{\authors}[1]{\noindent{\large #1}\vspace{5mm}}
\newcommand{\address}[1]{{\center{\noindent\small\itshape #1\vspace{0mm}}}}

\makeatletter
\def\section{\@startsection{section}{1}{\z@}{-3.25ex plus -1ex minus
             -.2ex}{1.5ex plus .2ex}{\normalfont\bfseries}}
\def\subsection{\@startsection{subsection}{1}{\z@}{-3.25ex plus -1ex
                minus -.2ex}{1.5ex plus .2ex}{\normalfont\itshape}}

\renewenvironment{thebibliography}[1]
         {\section*{References}\frenchspacing\small
          \begin{list}{[\arabic{enumi}]}
         {\usecounter{enumi}\parsep=2pt\topsep 0pt
         \settowidth{\labelwidth}{[#1]}
         \leftmargin=\labelwidth\advance\leftmargin\labelsep
         \rightmargin=0pt\itemsep=0pt\sloppy}}{\end{list}}

\@addtoreset{equation}{section}
\makeatother

\begin{document}

\begin{titlepage}
\begin{center}
\hspace*{\fill}{{\normalsize \begin{tabular}{l}
                              {\sf REF. TUW 02-11}\\
                              {\tt hep-th/0205153}
                              \end{tabular}   }}

\title{\vspace{5mm} Quantisation of $\theta$-expanded non-commutative QED}

\authors {{ Jesper M. Grimstrup$^{1}$, Raimar Wulkenhaar$^{2}$ }}
       
\address{$^{1}$  Institut f\"ur Theoretische Physik, 
Technische Universit\"at Wien \\
Wiedner Hauptstra\ss e 8--10, A-1040 Wien, Austria}

\address{$^{2}$ Max-Planck-Institut f\"ur Mathematik in den 
Naturwissenschaften\\
Inselstra\ss{}e 22--26, D-04103 Leipzig, Germany}

\footnotetext[1]{jesper@hep.itp.tuwien.ac.at, work supported by The
  Danish Research Agency.}

\footnotetext[2]{raimar.wulkenhaar@mis.mpg.de, Schloe\ss{}mann fellow}

\begin{minipage}{12cm}

\vspace{20mm}

{\it Abstract.}  We analyse two new versions of $\theta$-expanded
non-commutative quantum electrodynamics up to first order in $\theta$
and first loop order. In the first version we expand the bosonic
sector using the Seiberg-Witten map, leaving the fermions unexpanded.
In the second version we leave both bosons and fermions unexpanded.
The analysis shows that the Seiberg-Witten map is a field redefinition
at first order in $\theta$. However, at higher order in $\theta$ the
Seiberg-Witten map cannot be regarded as a field redefinition. We find
that the initial action of any $\theta$-expanded massless
non-commutative QED must include one extra term proportional to
$\theta$ which we identify by loop calculations.

\end{minipage}

\end{center}

\end{titlepage}

\section{Introduction}

Quantum field theory on non-commutative structures has received
increasing attention during the last years \cite{Seiberg:1999vs,
  Konechny:2000dp, Douglas:2001ba, Szabo:2001kg}. In almost all
articles on the subject a non-commutative structure
\begin{equation}
[\hat{x}^\m,\hat{x}^\n] = {\rm{i}}\q^{\m\n} \label{upsa}\; ,
\end{equation}
characterised by a constant non-commutativity parameter $\theta$ has
been considered, mainly due to the possibility of explicit
calculations. Some investigations of field theories involving a
non-constant $\theta$ have been performed \cite{Hawkins:1999gn,
  Steinacker:2001fj}. In any case, non-commutative configurations of
the form (\ref{upsa}) are to be seen in the spirit of deformation
quantisation \cite{Kontsevich:1997vb}, which form a subspace of all
possible non-commutative settings \cite{connes}. The algebra
(\ref{upsa}) with constant $\theta$ serves as the simplest possible
setting for a non-commutative quantum field theory.

Because the relation (\ref{upsa}) implies a non-locality of the
underlying space-time, the question of renormalisability and, as a
consequence hereof, the method of quantisation, is of central
interest.

The method mostly applied for quantisation takes the full,
non-expanded, non-local action as the source of Feynman rules. This
leads to damping phases which renders the non-planar sector UV-finite.
However, the entailed UV-IR mixing leads to new infrared divergences
\cite{Minwalla:1999px} which spoil renormalisability beyond first
loop order. A thorough analysis of the problem was given in
\cite{Chepelev:2000hm, Chepelev:1999tt}. In the special case of
non-commutative $\f^4$-theory methods of finite temperature quantum
field theory have been used to re-sum the perturbation series leading
to renormalisability \cite{Griguolo:2001ez}. This method might be
useful for non-commutative gauge theories as well. So far, however, the
problem of preserving the gauge symmetry has not been solved. Ideas
involving field redefinitions to overcome the IR-problems were
presented in \cite{Grimstrup:2002nr}. It is however unclear whether
this idea will prove fruitful because it transfers the problems to
higher $n$-point functions rather than removing them from the theory.
The only solution to the problem of quantising non-commutative gauge
theory known today is the introduction of supersymmetry
\cite{Ruiz:2000hu, Khoze:2000sy}, because the divergences present in
certain supersymmetric models are `soft' enough to render the UV-IR
mixing integrable \cite{Matusis:2000jf}.

An alternative method of quantisation was proposed in
\cite{Madore:2000en}: The non-commutativity structure $\theta$
apparently limits the choice of gauge group to that of a matrix
representation of a $U(N)$ gauge group. The choice of a more general
group automatically entails enveloping algebra valued fields rendering
the model seemingly meaningless. This problem is solved by expanding
the model in $\theta$ via the Seiberg-Witten map \cite{Seiberg:1999vs,
  Jurco:2001my, Bichl:2001yf}, which expresses the non-commutative
gauge field in terms of a commutative gauge field. The price paid is
however very high: $\theta$-expanded theories are truly power-counting
non-renormalisable and involve infinitely many vertices with an
arbitrary number of legs. In \cite{Bichl:2001nf} the purely bosonic
case was analysed for an Abelian group and in \cite{Bichl:2001cq} it
was shown that the photon self-energy is renormalisable to all orders.
In \cite{Wulkenhaar:2001sq}, however, $\theta$-expanded
non-commutative QED was proven non-renormalisable, putting this line
of quantisation to an apparent halt.

A central problem related to non-commutative field theories which
becomes urgently important in the second method of quantisation is the
choice of the action: In order to quantise a power-counting
non-renormalisable model one needs very strong symmetries. Symmetries
are to be found in the classical action, but the above scenario does
not give any criteria which dictate the form of the action, apart from
Lorentz and gauge symmetries and the demand that the limit
$\theta\rightarrow 0$ yields the commutative model. These bonds leaves
a huge space of possible actions. If the naive---initial---choice of
the commutative action equipped with appropriate star products should
prove renormalisable the non-commutativity must somehow yield
symmetries by itself. On the other hand, one could speculate whether
the quantisation procedure will by itself cast light on this question
by forcing extra terms to the initial naive choice and thereby lead us
to a suitable action.  In the light of recent phenomenological works
considering the $\theta$-expanded Standard Model via the
Seiberg-Witten map \cite{Calmet:2001na} we find it important to single
out the initial action by the behaviour of the Seiberg-Witten map on
quantum level.

In this paper we analyse two variations of the second method of
quantisation ($\theta$-expansion) up to first order in $\q$ and first
loop order. These two variations are closely related to the
$\q$-expanded non-commutative QED studied in \cite{Wulkenhaar:2001sq}
where both the bosonic and fermionic sectors were expanded via the
Seiberg-Witten map. Here we first consider $\theta$-expanded
non-commutative QED in the special case where the Seiberg-Witten map
is only applied to the bosons. The reason for this is speculative:
Whereas there are strong mathematical reasons for expanding the bosons
via the Seiberg-Witten map \cite{Jurco:2001my} there does not seem to
be urgent reasons for applying the Seiberg-Witten differential
equation to the fermions \cite{Grimstrup:2001ja}. Secondly, we
consider the case of $\theta$-expanded non-commutative QED without use
of the Seiberg-Witten map whatsoever. In connection to the above it is
natural to investigate this case in order to fully understand the role
of the Seiberg-Witten map.

What we find is partly encouraging: Up to unphysical field
redefinitions both studied settings coincide with the results of
\cite{Wulkenhaar:2001sq}. This means that the Seiberg-Witten map is
nothing but a field redefinition at first order in $\q$. However, we
find substantial evidence that this will not be the case at higher
orders in $\theta$, thus leaving a small window of hope open for
$\theta$-expanded models. It means, though, that the initial action of
{\it any} $\theta$-expanded model involving fermions must be extended
with at least one extra term which we identify. This extra term
suffices---in the massless case---for one-loop renormalisability
at first order in $\q$.

In the massive case two extra instabilities occur in the fermionic
sector, which cannot be removed in an obvious way. It thus appears
that we are cumulating evidence that {\it if} one persists on
considering $\theta$-expanded non-commutative Yang-Mills theory, then
the fermionic mass must be introduced via a Higgs mechanism.

Let us mention that our initial motivation for studying
$\theta$-expanded field theories without using the Seiberg-Witten map
was first of all the wish to obtain $\theta$-graded symmetries which
could fix the $\theta$-structure of the action on the quantum level.
It turned out, however, that this does not work because we loose at
the same time the initial `flat' gauge symmetry. One could speculate
if other symmetries present in the original $\theta$-non-expanded
action (we think e.g.\ of symmetries of the spectral action
\cite{Chamseddine:1996zu}, see also the discussion in
\cite{Wulkenhaar:2001sq}) could provide useful $\theta$-graded
symmetries of the expanded action. An interesting example is
supersymmetry which indeed yields such a $\theta$-graded symmetry
\cite{Putz:2002ib}.

The paper is organised as follows: In section~\ref{sec2} we introduce
non-commutative Yang-Mills theory. In section~\ref{sec3} we expand the
action using the Seiberg-Witten map in the bosonic sector only. The
quantisation is studied at first loop order. In section~\ref{sec4} we
repeat the analysis for the action expanded in $\theta$ without the
Seiberg-Witten map. In section~\ref{sec5} we analyse general changes
of variables in the path-integral and, finally, in section~\ref{sec6}
we give a conclusion.

\section{Non-commutative Yang-Mills theory}
\label{sec2}

On the space of rapidly decreasing functions on $\mathbb{R}^4$ one
introduces a deformed product
\begin{equation}
(f\star g)(x)=\int \frac{d^4 k}{(2\pi)^4}
\int \frac{d^4 p}{(2\pi)^4} \,\mathrm{e}^{-\mathrm{i}
(k_\mu +p_\mu )x^\mu } \,
\mathrm{e}^{-\frac{\mathrm{i}}{2}\theta^{\mu\nu} k_\mu p_\nu }
\tilde{f}(k)\,\tilde{g}(p)\;,
\label{star}
\end{equation}
where $f,g$ are rapidly decreasing functions on the manifold and
$\tilde{f},\tilde{g}$ their Fourier transforms\footnote{Our Fourier
  conventions are $f(x)=\int \frac{d^4k}{(2\pi)^4} \;
  \mathrm{e}^{-\mathrm{i} k\cdot x} \, \tilde{f}(k)$ and
  $\tilde{f}(k)=\int d^4x\; \mathrm{e}^{\mathrm{i} k\cdot x} \,
  f(x)$.}. The $\star$-product (\ref{star}) is associative and
non-commutative and yields for the coordinates $x^\mu$ belonging to
the multiplier algebra the commutator
\begin{equation} 
[x^\m,x^\n]_{\star}\equiv x^\m \star x^\n -
x^\n \star x^\m = {\rm{i}}\q^{\m\n}\;.
\label{st2}
\end{equation}

We consider the action of non-commutative Yang-Mills (NCYM) theory,
including fermions, given by
\begin{align}
\hat{\Sigma}_{cl} &=\int d^{4}x\,\Big( -\frac{1}{4g^2} \hat{F}_{\mu\nu} 
\star \hat{F}^{\mu\nu}
+ \Hat{\Bar{\psi}} \star \mathrm{i} \gamma^\mu \hat{D}_\mu \hat{\psi} 
-m \Hat{\Bar{\psi}} \star \hat{\psi} \Big) \;,
\label{theaction}
\end{align}
with
\begin{align}
\hat{F}_{\m\n} &=  \partial_{\m}\hat{A}_{\n}-\partial_{\n}\hat{A}_{\m}
-{\rm{i}}\big[\hat{A}_{\m}, \hat{A}_{\n}]_\star \; , \nonumber
\\
\hat{D}_{\m}\hat{\j} & =   \partial_{\m}\hat{\j}
-\mathrm{i} \hat{A}_{\m}\star \hat{\j}\;.
\nonumber
\end{align}
Notice that this is a {\it non-local} field theory. The action
(\ref{theaction}) is invariant with respect to an infinitesimal gauge
transformation 
\begin{align}
\d_{\hat{\Ld}}\hat{A}_{\m} & = \hat{D}^{\mathit{adj}}_{\m}\hat{\Ld} 
\equiv \partial_{\m}\hat{\Ld} -{\rm{i}}[\hat{A}_\m,\hat{\Ld}]_{\star}\;, 
\nonumber 
\\
\d_{\hat{\Ld}}\hat{\j} & = \mathrm{i} \hat{\Ld}\star\hat{\j}\;, 
\nonumber
\\
\d_{\hat{\Ld}}\Hat{\Bar{\j}} & = -\mathrm{i}
\Hat{\Bar{\j}}\star\hat{\Ld}\;.
\label{BRST}
\end{align}
Usually this gauge symmetry is fixed via a gauge-fixing term,
introducing ghost $\hat{c}$, anti-ghost $\Hat{\Bar{c}}$ and multiplier
field $\Hat{B}$,
\begin{align}
\hat{\Sigma}_{gf}&= \int d^4x\;\hat{s}\left(\Hat{\Bar{c}}\star \partial^\mu
  \hat{A}_\mu +\frac{\alpha}{2}\,\Hat{\Bar{c}}\star\hat{B} \right)\;, 
\label{hSgf}
\end{align}
where we define the non-commutative BRST-transformations by
\begin{align}
\hat{s} \hat{A}_{\m} &=  \hat{D}^{\mathit{adj}}_{\m}\hat{c} \;,& 
\hat{s} \hat{c} &= \mathrm{i}\hat{c}\star \hat{c} \;,& 
\nonumber\\
\hat{s} \hat{\j} &= {\rm{i}}\hat{c}\star\hat{\j} \;, &
\hat{s} \Hat{\Bar{\j}} &= - {\rm{i}}\Hat{\Bar{\j}} \star \hat{c} \;, &
\nonumber\\
\hat{s} \Hat{\Bar{c}} &= \hat{B} \;,& 
\hat{s} \hat{B} &= 0 \;.&
\end{align}
Finally, we couple the non-linear BRST-transformations to external
fields $(\hat{\Bar{\r}},\hat{\r},\hat{\s}^\m,\hat{\k})$ by introducing
an extra term to the action, 
\begin{align} 
\hat{\S}_{ext} &= \int d^4x\;\Big( \hat{\Bar{\r}} \big( \hat{s}\hat{\j}\big) 
+ \big(\hat{s} \Hat{\Bar{\j}}\big)\hat{\r} 
+ \hat{\s}^\m\big( \hat{s} \hat{A}_\m \big) 
+ \hat{\k}\big( \hat{s} \hat{c}\big) \Big) \;,
\end{align}
and define 
\begin{align}
\hat{s}\hat{\r}&=0\;, &
\hat{s}\Hat{\Bar{\r}}&=0\;, &
\hat{s}\hat{\s}^\m&=0\;, & 
\hat{s}\hat{\k} &= 0\;.  
\end{align}
The full tree level generating functional for 1PI graphs reads 
\begin{align}
\hat{\Gamma}^{(0)} =\left( \hat{\Sigma}_{cl} + \hat{\Sigma}_{ext} 
+ \hat{\Sigma}_{gf}\right) \;.\label{fullaction} 
\end{align}
The nilpotency of $\hat{s}$ allows us to write down the Slavnov-Taylor
identity expressing the BRST-symmetry: 
\begin{align}
\mathcal{S}\left(\hat{\Gamma}^{(0)}\right)&= 0\;, \label{sla} 
\\
\mathcal{S}\left(\Gamma\right) &= \int d^4x\;\bigg(
\frac{\delta\Gamma}{\delta \hat{\sigma}^\mu}
  \frac{\delta\Gamma}{\delta \hat{A}_\mu} 
+\frac{\delta\Gamma}{\delta \hat{\kappa}}
  \frac{\delta\Gamma}{\delta \hat{c}} 
+\frac{\delta\Gamma}{\delta \hat{\rho}} 
  \frac{\delta\Gamma}{\delta \Hat{\Bar{\psi}}}
+\frac{\delta\Gamma}{\delta\hat{\psi} } 
  \frac{\delta\Gamma}{\delta \hat{\bar{\rho}}}
+ \hat{B}\frac{\delta\Gamma}{\delta \Hat{\Bar{c}}} \bigg)\;.
\label{SlA}
\end{align}

\section{Expanding the action. [Case I]}
\label{sec3}

In this section we expand the action of NCYM theory in $\theta$ using
the Seiberg-Witten differential equation in the bosonic sector. In
contrast to the analysis performed in \cite{Wulkenhaar:2001sq} on $\q$-expanded QED \cite{Bichl:2001gu} the
fermions are not expanded. This entails a picture where the gauge
symmetry is `flat' (not $\theta$-graded) in the bosonic sector and
`tilted' ($\theta$-graded) in the fermionic sector. Performing the
loop calculations we show that the Slavnov-Taylor identities are still
valid on quantum level. Also, we find that the model characterised by
the classical action (\ref{theaction}) is instable. Remarkable enough,
the bosonic sector is stable at first order in $\theta$. These results
are up to field redefinitions identical to the results found in
\cite{Wulkenhaar:2001sq}.

\subsection{Classical analysis}

The expansion of the action (\ref{fullaction}) is performed according to
\begin{align}
\big(f \star g\big)(x) &= f(x)g(x) +
\frac{{\rm{i}}}{2}\theta^{\a\b}\partial_\a f(x)\partial_\b g(x)
+ \mathcal{O}(\q^{2}) \;,\nonumber
\\
\hat{A}_\m &= A_{\m}- \frac{1}{2}\q^{\r\s}
A_{\r}\left(\partial_{\s}A_{\m}+F_{\s\m}\right)
+ \mathcal{O}(\q^{2})\;,\nonumber
\\
\hat{\Phi} &= \Phi\;, \qquad \forall\;\hat{\Phi} \in\{\hat{\psi},
\hat{\bar{\psi}},\hat{c},\hat{\bar{c}},\hat{B},\hat{\bar{\rho}},
\hat{\rho},\hat{\sigma}^\m,\hat{\kappa} \}\;,
\label{expa}
\end{align}
where the gauge field is expanded according to the Seiberg-Witten
differential equation \cite{Seiberg:1999vs}. This leads to the
expanded action
\begin{align}
\Sigma^{\{n\}}_{\theta\textit{-exp}}
&= \sum_{i=0}^n\Sigma^{(i)} \;,
\end{align}
which up to first order in $\theta$ (which we are interested in from
now on) reads
\begin{align}
\Sigma^{(0)}_{cl} &= \int d^4x
\left(-\frac{1}{4 g^2} F_{\mu\nu}F^{\mu\nu}
+ \bar{\psi} \left( \mathrm{i} \gamma^\mu D_\mu -m \right) \psi \right)\;,
\\
\Sigma^{(1)}_{cl} &= \int d^4x
\bigg( \frac{1}{8 g^2}\theta^{\alpha\beta} F_{\alpha\beta} F_{\mu\nu} 
F^{\mu\nu}
-\frac{1}{2g^2 }\theta^{\alpha\beta}F_{\mu\alpha}F_{\nu\beta} 
F^{\mu\nu}
\nonumber
\\
& \qquad\quad + \frac{\mathrm{i}}{2} \theta^{\alpha\beta} 
\bar{\psi} \gamma^\mu \partial_\alpha A_\mu \partial_\beta \psi 
-\theta^{\alpha\beta} \bar{\psi} \gamma^\mu A_\alpha 
\partial_\beta A_\mu \psi 
+\frac{1}{2} \theta^{\alpha\beta} \bar{\psi} \gamma^\mu A_\alpha 
\partial_\mu A_\beta \psi \bigg)\;,
\\
\Sigma^{(0)}_{gf} &= \int d^4x \Big( B \partial^\mu A_\mu +
\frac{\alpha}{2} B B - \bar{c} \partial^\mu \partial_\mu c \Big)\;,
\label{Sgf0}
\\
\Sigma^{(\geq 1)}_{gf} &= 0\;. \label{Sgf1}
\end{align}
We choose the `linear gauge-fixing' in the sense of
\cite{Bichl:2001nf} applied after the Seiberg-Witten map (\ref{expa}).
The $\theta$-expansion of (\ref{hSgf}) leads to the `non-linear
gauge-fixing' which is different\footnote{In [Case I] we apply the
  linear gauge, which is possible because the BRST-symmetry is `flat'
  in the bosonic sector---in perfect analogy to \cite{Bichl:2001nf}.
  In [Case II] (see section \ref{sec4}) we use (a variation of) the
  non-linear gauge, because in [Case II] we have a $\theta$-graded
  BRST symmetry in the bosonic sector, leaving no room for a linear
  gauge.  Since we have shown that the choice of linear/non-linear
  gauge leaves loop calculations invariant \cite{Bichl:2001nf} this is
  justified.}. We expand the BRST transformations (\ref{BRST})
according to (\ref{expa}),
\begin{align}
\hat{s} &= \sum_{i} s^{(i)}\;, \nonumber
\end{align}
which to first order in $\theta$ gives
\begin{align}
s^{(0)}A_\m &= \partial_\m c \;,&  s^{(0)}c &= 0\;, \nonumber
\\
s^{(0)}\j &= {\rm{i}} c \j \;, &
s^{(0)}\bar{\j} &= -{\rm{i}} \bar{\j}c \;,  \nonumber
\\
s^{(1)}A_\m &= 0 \;, &  s^{(1)}c &= 0 \;, \nonumber
\\
s^{(1)}\j &=  -\frac{{\rm{i}}}{2}\q^{\a\b}A_\a\partial_\b c \j 
- \frac{1}{2}\q^{\a\b}\partial_\a c\partial_\b\j \;,&
s^{(1)}\bar{\j} &= \frac{{\rm{i}}}{2}\q^{\a\b}\bar{\j}A_\a\partial_\b c  
+\frac{1}{2}\q^{\a\b} \partial_\a\bar{\j}\partial_\b c\;. \label{bb}
\end{align}
Here we have used the Seiberg-Witten expansion of the non-commutative
gauge parameter \cite{Seiberg:1999vs}
\begin{align}
\hat{\Lambda}&=\lambda-\frac{1}{2}\theta^{\alpha\beta} A_{\alpha}
\partial_{\beta}\lambda +\mathcal{O}(\q^2)\;,
\end{align}
replacing $\lambda$ by $c$ \cite{Bichl:2001nf}, to obtain the
non-commutative gauge transformation of the fermions in terms of the
commutative gauge parameter. Notice that only the BRST transformations
of the fermions (\ref{bb}) are $\theta$-graded. The application of the
Seiberg-Witten map in the bosonic sector `flattens' out their
BRST-transformations.  The total $\theta$-expanded action is invariant
under non-commutative BRST transformations
\begin{equation}
s \sum_{i}\S^{(i)}   =0\;,
\label{BRST-grad}
\end{equation}
leading to a tower of symmetries
\begin{align}
s^{(0)}\S^{(0)} &= 0\;, \label{nr1}
\\*
s^{(1)}\S^{(0)}+s^{(0)}\S^{(1)}  &= 0\;, \label{nr2}
\\*
s^{(2)}\S^{(0)}+s^{(1)}\S^{(1)} +s^{(0)}\S^{(2)} &= 
0\;,\label{nr3} 
\\*
&\vdots \nonumber
\end{align}
where (\ref{nr1}) is simply the BRST invariance of the commutative
theory. 

\subsection{Slavnov-Taylor identity}

Loop corrections do not preserve the BRST symmetry in the form
(\ref{BRST-grad}). The solution of this problem is to couple the
non-linear BRST transformations to external fields, 
\begin{align}
\S_{ext}^{(n)} &= \int d^4x\;\Big( \s^\m \left( s^{(n)}A_\m\right) 
+ \k\left( s^{(n)}c\right) + \Bar{\r} \left( s^{(n)} \j\right) 
+ \left(s^{(n)}\Bar{\j}\right)\r \Big)\,.
\label{EXTER}
\end{align}
Defining the full tree level generating functional for 1PI graphs 
to $n^{\mathrm{th}}$ order in $\theta$ by
\begin{align}
\Gamma^{(n,0)} =\left( \Sigma^{(n)}_{\theta\textit{-exp}} 
+ \Sigma_{ext}^{(n)} + \Sigma_{gf}^{(n)}
\right)\;,
\label{XX}
\end{align}
the Slavnov-Taylor identity expresses the whole set of
BRST invariances (\ref{nr1})--(\ref{nr3}) up to $n^{\mathrm{th}}$
order in $\theta$,
\begin{align} 
\Big(\mathcal{S}\G\Big)^{(n)} &=0\;, 
\label{sla1}
\end{align}
where the Slavnov-Taylor operator is defined by (\ref{SlA}) (without
the hat over the fields). In momentum space we have
\begin{align}
0 &=\int \!\!\frac{d^4 k}{(2\pi)^4} \Big( 
\frac{\delta \Gamma}{\delta \sigma^\mu(k)}  
\frac{\delta \Gamma}{\delta A_\mu(-k)}  
+ \frac{\delta \Gamma}{\delta \rho(k)}  
\frac{\delta \Gamma}{\delta \bar{\psi}(-k)}  
+ \frac{\delta \Gamma}{\delta \psi(k)}  
\frac{\delta \Gamma}{\delta \bar{\rho}(-k)}  
+ \frac{\delta \Gamma}{\delta \kappa(k)}  
\frac{\delta \Gamma}{\delta c(-k)}  
\nonumber
\\
& \quad \qquad + B(k) \frac{\delta \Gamma}{\delta \bar{c}(k)} \Big)\;.
\label{ST}
\end{align}
Functional derivation of (\ref{ST}) with respect to the fields
$\{A_\mu,c,\psi,\bar{\psi},\bar{c},B\}$ in momentum space,
followed by putting the fields to zero, leads to various forms of the
Slavnov-Taylor identity for 1PI Green's functions
\begin{align}
(2\pi)^4 \delta(p_1{+}\dots{+}p_N)\; 
\Gamma_{\Phi_1\dots\Phi_N}(p_1,\dots,p_N) :=\frac{\delta^N
  \Gamma}{\delta \Phi_N(p_N) \dots \delta \Phi_1(p_1)} 
\Big|_{\Phi_i =0} \;.
\end{align}
These Green's functions $\Gamma_{\Phi_1\dots\Phi_N}(p_1,\dots,p_N) =
\sum_{n,\ell \geq 0}
\Gamma^{(n,\ell)}_{\Phi_1\dots\Phi_N}(p_1,\dots,p_N)$ carry a bidegree
$(n,\ell)$ where $n$ is the number of factors of $\theta$ and $\ell$
the number of loops. For our purpose the most important Slavnov-Taylor
identities derived from (\ref{ST}) are the following:
\begin{align}
0 &= \sum_{\ell'=0}^\ell \sum_{n'=0}^n \Big(
\Gamma^{(n',\ell')}_{\mu;\sigma c}(q{+}r,p) 
\Gamma^{(n-n',\ell-\ell')\mu}_{A\bar{\psi}\psi}(p,q,r) 
\nonumber
\\*
& \qquad\quad 
+ \Gamma^{(n',\ell')}_{\bar{\psi} c\rho }(q,p,r) 
\Gamma^{(n-n',\ell-\ell')}_{\bar{\psi}\psi}(p{+}q,r) 
+ \Gamma^{(n-n',\ell-\ell')}_{\bar{\psi}\psi}(q,p{+}r) 
\Gamma^{(n',\ell')}_{\bar{\rho} c\psi}(q,p,r)\Big)\;,
\label{STAjj}
\\
0 &= \sum_{\ell'=0}^\ell \sum_{n'=0}^n \Big(
\Gamma^{(n',\ell')}_{\mu;\sigma c}(q{+}r{+}s,p) 
\Gamma^{(n-n',\ell-\ell')\mu\nu}_{AA\bar{\psi}\psi}(p,q,r,s) 
\nonumber
\\*
& \qquad\quad 
+ \Gamma^{(n',\ell')\nu}_{\mu;A\sigma c}(q,r{+}s,p) 
\Gamma^{(n-n',\ell-\ell')\mu}_{A\bar{\psi}\psi}(p{+}q,r,s) 
\nonumber
\\*
& \qquad\quad 
+ \Gamma^{(n',\ell')\nu}_{A\bar{\psi} c\rho }(q,r,p,s) 
\Gamma^{(n-n',\ell-\ell')}_{\bar{\psi}\psi}(p{+}q{+}r,s) 
+ \Gamma^{(n-n',\ell-\ell')}_{\bar{\psi}\psi}(r,p{+}q{+}s) 
\Gamma^{(n',\ell')\nu}_{A\bar{\rho} c\psi}(q,r,p,s)
\nonumber
\\*
&
\qquad\quad + \Gamma^{(n',\ell')}_{\bar{\psi} c\rho }(r,p,q{+}s) 
\Gamma^{(n-n',\ell-\ell')\nu}_{A\bar{\psi}\psi}(q,p{+}r,s) 
+ \Gamma^{(n-n',\ell-\ell')\nu}_{A\bar{\psi}\psi}(q,r,p{+}s) 
\Gamma^{(n',\ell')}_{\bar{\rho} c\psi}(q{+}r,p,s)\Big)\;,
\label{STAAjj}
\\
0 &= \sum_{\ell'=0}^\ell \sum_{n'=0}^n 
\Gamma^{(n',\ell')}_{\mu;\sigma c}(q,p) 
\Gamma^{(n-n',\ell-\ell')\mu\nu}_{AA}(p,q)\;,
\label{STAA}
\\
0 &= \sum_{\ell'=0}^\ell \sum_{n'=0}^n \Big(
\Gamma^{(n',\ell')\nu}_{\mu;A\sigma c}(q,r,p) 
\Gamma^{(n-n',\ell-\ell')\mu\rho}_{AA}(p{+}q,r)
+ \Gamma^{(n',\ell')\rho}_{\mu;A\sigma c}(r,q,p) 
\Gamma^{(n-n',\ell-\ell')\mu\nu}_{AA}(p{+}r,q)
\nonumber
\\*
& \qquad\quad 
+ \Gamma^{(n',\ell')}_{\mu;\sigma c}(q{+}r,p) 
\Gamma^{(n-n',\ell-\ell')\mu\nu\rho}_{AAA}(p,q,r)
\Big)\;,
\label{STAAA}
\\
0 &=\sum_{\ell'=0}^\ell \sum_{n'=0}^n \Big( 
\Gamma^{(n',\ell')}_{\mu;\sigma c}(q{+}r{+}s,p)
\Gamma^{(n-n',\ell-\ell')\mu}_{A\bar{\psi}c\rho}(p,r,q,s) 
-\Gamma^{(n',\ell')}_{\mu;\sigma c}(p{+}r{+}s,q)
\Gamma^{(n-n',\ell-\ell')\mu}_{A\bar{\psi}c\rho}(q,r,p,s) 
\nonumber
\\*
& 
\qquad\quad 
+ \Gamma^{(n',\ell')}_{\bar{\psi}c\rho}(r,p,q{+}s)
\Gamma^{(n-n',\ell-\ell')}_{\bar{\psi}c\rho}(p{+}r,q,s) 
- \Gamma^{(n',\ell')}_{\bar{\psi}c\rho}(r,q,p{+}s)
\Gamma^{(n-n',\ell-\ell')}_{\bar{\psi}c\rho}(q{+}r,p,s) 
\nonumber
\\*
& 
\qquad\quad
+ \Gamma^{(n',\ell')}_{\kappa cc}(r{+}s,p,q)
\Gamma^{(n-n',\ell-\ell')}_{\bar{\psi}c\rho}(r,p{+}q,s)\Big)\;,
\label{STAjcr}
\\
0 &=\sum_{\ell'=0}^\ell \sum_{n'=0}^n \Big( 
\Gamma^{(n',\ell')}_{\mu;\sigma c}(q{+}r{+}s,p)
\Gamma^{(n-n',\ell-\ell')\mu}_{A\bar{\rho}c\psi}(p,r,q,s) 
-\Gamma^{(n',\ell')}_{\mu;\sigma c}(p{+}r{+}s,q)
\Gamma^{(n-n',\ell-\ell')\mu}_{A\bar{\rho}c\psi}(q,r,p,s) 
\nonumber
\\*
& 
\qquad\quad 
- \Gamma^{(n',\ell')}_{\bar{\rho}c\psi}(r,p,q{+}s)
\Gamma^{(n-n',\ell-\ell')}_{\bar{\rho}c\psi}(p{+}r,q,s) 
+ \Gamma^{(n',\ell')}_{\bar{\rho}c\psi}(r,q,p{+}s)
\Gamma^{(n-n',\ell-\ell')}_{\bar{\rho}c\psi}(q{+}r,p,s) 
\nonumber
\\*
& 
\qquad\quad
+ \Gamma^{(n',\ell')}_{\kappa cc}(r{+}s,p,q)
\Gamma^{(n-n',\ell-\ell')}_{\bar{\rho}c\psi}(r,p{+}q,s)\Big)\;.
\label{STArcj}
\end{align}
For $n{=}0$ we recover ordinary QED, where additionally $\ell'{=}0$
because there are no loops involving external fields. 

The above Slavnov-Taylor identities can be verified
on a formal level of divergent integrals and hold for renormalised
Green's functions when using an invariant regularisation scheme.
However, in contrast to the commutative world, the Slavnov-Taylor
identities are in presence of $\theta$ not strong enough to preserve
the form of the action at higher loop order. 

\subsection{The tree-level Green's functions}

To be explicit, the non-vanishing tree-level Green's functions of our
model are at order $n{=}0$ in $\theta$ given by
\begin{align}
\Gamma^{(0,0)}_{\bar{\psi}\psi}(q,p) &= \gamma^\mu p_\mu -m\;,
&
\Gamma^{(0,0)\mu}_{A\bar{\psi}\psi}(p,q,r) &= \gamma^\mu \;,
\nonumber
\\
\Gamma^{(0,0)\mu\nu}_{AA}(p,q) &= -\frac{1}{g^2}\Big(p^2 g^{\mu\nu} 
- p^\mu p^\nu\Big) \;,
&
\Gamma^{(0,0)\mu}_{AB}(p,q) &= -\mathrm{i} p^\mu \;,
\nonumber
\\
\Gamma^{(0,0)}_{BB}(p,q) &= \alpha\;,
&
\Gamma^{(0,0)}_{\bar{c}c}(q,p) &= p^2\;,
\nonumber
\\
\Gamma^{(0,0)}_{\bar{\rho}c\psi}(q,p,r) &= \mathrm{i}\;,
&
\Gamma^{(0,0)}_{\bar{\psi}c\rho}(q,p,r) &= -\mathrm{i}\;,
\nonumber
\\
\Gamma^{(0,0)}_{\mu;\sigma c}(q,p) &= -\mathrm{i}p_\mu\;.
\label{G00sc}
\end{align}
It is straightforward to check the tree-level $(\ell{=}0)$
Slavnov-Taylor identities (\ref{STAjj})--(\ref{STArcj}) for $n{=}0$. The
propagators are the bilinear parts of the tree-level generating
functional of connected Green's functions:
\begin{align}
\Delta^{\bar{\psi}\psi}(q,p) &= - \frac{\gamma^\mu p_\mu +m}{
p^2-m^2+\mathrm{i}\epsilon} \;,
&
\Delta^{AA}_{\mu\nu}(p,q) &= \frac{g^2}{p^2+\mathrm{i}\epsilon} 
\Big( g_{\mu\nu} -\Big(1-\frac{\alpha}{g^2}\Big) 
\frac{p_\mu p_\nu}{p^2+\mathrm{i}\epsilon} \Big)\;,
\nonumber
\\*
\Delta^{AB}_{\mu}(p,q) &= -\frac{\mathrm{i}
  p_\mu}{p^2+\mathrm{i}\epsilon} 
&
\Delta^{\bar{c}c}(q,p) &= -\frac{1}{p^2+\mathrm{i}\epsilon} \;.
\end{align}
At order $n{=}1$ in $\theta$ we have the following tree-level 1PI Green's
functions: 
\begin{align}
\Gamma^{(1,0)\mu}_{A\bar{\psi}\psi}(p,q,r) &= -\frac{\mathrm{i}}{2}
\theta^{\alpha\beta} p_\alpha r_\beta \gamma^\mu\;,
\\
\Gamma^{(1,0)\mu\nu}_{AA\bar{\psi}\psi}(p,q,r,s) &= 
 \mathrm{i} \theta^{\mu\beta} q_\beta \gamma^\nu 
+ \mathrm{i} \theta^{\nu\beta} p_\beta \gamma^\mu 
-\frac{\mathrm{i}}{2} \theta^{\mu\nu} (q_\rho-p_\rho) \gamma^\rho \;,
\\
\Gamma^{(1,0)\mu\nu\rho}_{AAA}(p,q,r) 
& = \frac{1}{g^2}
\mathrm{i} \theta_{\alpha\beta} \Big(
g^{\alpha\mu}g^{\beta\nu}\big( (qr)p^{\rho} - (pr)q^{\rho} 
\big)
+g^{\alpha\nu}g^{\beta\rho}\big( (rp)q^{\mu} - (qp)r^{\mu} \big)
\nonumber
\\*
&
+g^{\alpha\rho}g^{\beta\mu}\big( (pq)r^{\nu} - (rq)p^{\nu} \big) 
\nonumber
\\
&+ g^{\alpha\mu} \big( 
( g^{\nu\rho} (pq) - p^\nu q^\rho ) r^{\beta}
+( g^{\nu\rho} (rp) - r^\nu p^\rho ) q^{\beta}\big) 
\nonumber
\\
&+ g^{\alpha\nu} \big( 
( g^{\rho\mu} (qr) - q^\rho r^\mu ) p^{\beta}
+( g^{\rho\mu} (pq) - p^\rho q^\mu ) r^{\beta}\big) 
\nonumber
\\*
&+ g^{\alpha\rho} \big( 
( g^{\mu\nu} (rp) - r^\mu p^\nu ) q^{\beta}
+( g^{\mu\nu} (qr) - q^\mu r^\nu ) p^{\beta}\big) 
\nonumber
\\
&+ g^{\mu\nu}\big(p^{\rho} q^{\alpha} r^{\beta} 
+ q^{\rho} p^{\alpha} r^{\beta} \big)
 + g^{\nu\rho}\big(q^{\mu} r^{\alpha} p^{\beta} 
+ r^{\mu} q^{\alpha} p^{\beta} \big)
+ g^{\rho\mu}\big(r^{\nu} p^{\alpha} q^{\beta} 
+ p^{\nu} r^{\alpha} q^{\beta} \big)
\nonumber
\\*
&-g^{\alpha\mu} ( g^{\nu\rho} (rq) - r^\nu q^\rho ) p^{\beta}
- g^{\alpha\nu}  ( g^{\rho\mu} (pr) - p^\rho r^\mu ) q^{\beta}
\nonumber
\\*
&- g^{\alpha\rho} ( g^{\mu\nu} (qp) - q^\mu p^\nu ) r^{\beta} \Big),
\\
\Gamma^{(1,0)}_{\bar{\rho}c\psi}(q,p,r) &= \frac{1}{2}
  \theta^{\alpha\beta} p_\alpha r_\beta~,
\\
\Gamma^{(1,0)}_{\bar{\psi}c\rho}(q,p,r) &= \frac{1}{2}
  \theta^{\alpha\beta} p_\alpha q_\beta~,
\\
\Gamma^{(1,0)\nu}_{A\bar{\rho}c\psi}(q,r,p,s) &= -\frac{1}{2}
  \theta^{\nu\beta} p_\beta~,
\\
\Gamma^{(1,0)\nu}_{A\bar{\psi}c\rho}(q,r,p,s) &= \frac{1}{2}
  \theta^{\nu\beta} p_\beta~.
\end{align}
It is straightforward to check the tree-level $(\ell{=}0)$ Slavnov-Taylor
identities (\ref{STAjj})--(\ref{STArcj}) for $n{=}1$.

\subsection{One-loop computation}

Using analytic regularization we compute the one-loop divergent
Green's functions up to first order in $\theta$. We choose the Feynman
gauge $\alpha=g^2$. At order $n{=}0$ in $\theta$ we find
\begin{align}
\Gamma^{(0,1)}_{\bar{\psi}\psi}(q,p) &= \frac{\hbar
  g^2}{(4\pi)^2 \varepsilon} 
\Big( \frac{1}{2} N_\psi + 3 m \frac{\partial}{\partial m}\Big)
\Gamma^{(0,0)}_{\bar{\psi}\psi}(q,p) \;,
\label{G01jj}
\\
\Gamma^{(0,1)\mu}_{A\bar{\psi}\psi}(p,q,r) &= \frac{\hbar
  g^2}{(4\pi)^2 \varepsilon}\Big( \frac{1}{2} N_\psi + 0 N_A \Big)
\Gamma^{(0,0)\mu}_{A\bar{\psi}\psi}(p,q,r) \;,
\label{G01Ajj}
\\
\Gamma^{(0,1)\mu\nu}_{AA}(p,q) &= \frac{\hbar
  g^2}{(4\pi)^2 \varepsilon}\Big( -\frac{4}{3} g^2
\frac{\partial}{\partial g^2} + 0 N_A \Big)
\Gamma^{(0,0)\mu\nu}_{AA}(p,q) \;,
\label{G01AA}
\end{align}
where $N_\psi$ and $N_A$ are the counting operators of electrons
$\bar{\psi},\psi$ and photons $A_\mu$, respectively. There are no
divergences in graphs involving
$c,\bar{c},B,\sigma^\mu,\rho,\bar{\rho}$ at order 0 in $\theta$ so
that $\rho,\bar{\rho}$ must receive a wave function renormalisation
$-\frac{1}{2} \frac{\hbar g^2}{(4\pi)^2 \varepsilon} N_\rho$ in order
to compensate the wave function renormalisation of $\psi,\bar{\psi}$.
The result (\ref{G01jj})--(\ref{G01AA}) means that at order 0 in
$\theta$ all one-loop divergences can be removed by a redefinition of
the electron wave function, the electron mass and the coupling
constant.

At order $n{=}1$ in $\theta$ we find
\begin{align}
\Gamma^{(1,1)}_{\bar{\psi}\psi}(q,p) &= 0 \;,
\label{G11jj}
\\
\Gamma^{(1,1)\mu}_{A\bar{\psi}\psi}(p,q,r) &= \frac{\hbar
  g^2}{(4\pi)^2 \varepsilon} 
\Big(\Big( \frac{1}{2} N_\psi + 0 N_A
\Big)\Gamma^{(1,0)\mu}_{A\bar{\psi}\psi}(p,q,r)
\nonumber
\\*
& 
+ \mathrm{i}\theta^{\alpha\beta} \Big(
\frac{1}{2} (p_\nu r_\beta - r_\nu p_\beta) \delta^\mu_\alpha \gamma^\nu
-\frac{1}{4} m \delta^\mu_\alpha (2 r^\nu + p^\nu)\gamma_{\beta\nu} 
\nonumber
\\*
& 
- \frac{3}{4} (p^2 \delta^\mu_\nu - p^\mu  p_\nu)
\gamma^\nu_{~\alpha\beta} 
- \frac{3}{2} p_\nu p_\beta \gamma^{\mu\nu}_{~~\alpha} 
+ \frac{15}{4} m \delta^\mu_\alpha p_\beta 
\Big)\Big)\;,
\label{G11Ajj}
\\
\Gamma^{(1,1)\mu\nu}_{AA\bar{\psi}\psi}(p,q,r,s) &=\dots\;,
\label{G11AAjj}
\\
\Gamma^{(1,1)\mu\nu\rho}_{AAA\bar{\psi}\psi}(p,q,r,s,t) &=\dots\;,
\label{G11AAAjj}
\\
\Gamma^{(1,1)}_{\bar{\psi}\psi\bar{\psi}\psi} (p,q;r,s)&=  \frac{\hbar
  g^2}{(4\pi)^2 \varepsilon} \mathrm{i}\theta^{\alpha\beta} 
\Big( \frac{3}{4} g^2 \gamma^\mu \otimes
\gamma_{\mu\alpha\beta} 
\Big)\;,
\label{G11jjjj}
\\
\Gamma^{(1,1)\mu\nu}_{AA}(p,q) &= 0 \;,
\label{G11AA}
\\
\Gamma^{(1,1)\mu\nu\rho}_{AAA}(p,q,r) &= 
\frac{\hbar g^2}{(4\pi)^2 \varepsilon}
\Big(-\frac{4}{3} g^2 \frac{\partial}{\partial g^2}
\Big)\Gamma^{(1,0)\mu\nu\rho}_{AAA}(p,q,r)\;,
\label{G11AAA}
\\
\Gamma^{(1,1)\mu_1\dots \mu_N}_{A\dots A}(p_1,\dots p_N) &= \dots \,, 
\qquad N\in\{4,5,6\}\;.
\label{G11AN}
\end{align}
We did not compute the divergent Green's functions
$\Gamma^{(1,1)\mu\nu}_{AA\bar{\psi}\psi}(p,q,r,s)$,
$\Gamma^{(1,1)\mu\nu\rho}_{AAA\bar{\psi}\psi}(p,q,r,s,t)$ and 
$\Gamma^{(1,1)\mu_1\dots \mu_N}_{A\dots A}(p_1,\dots p_N)$ for
$N\in\{4,5,6\}$, because
they do not give new information for the discussion (see
footnote~\ref{AA} below). The graphs to compute for the Green's
functions (\ref{G11jj}), (\ref{G11Ajj}) and
(\ref{G11jjjj})--(\ref{G11AAA}) 
are exactly the same as those
given in \cite{Wulkenhaar:2001sq}, only the Feynman rules are
different. There is no need to print these graphs again. However,
there are now divergent graphs involving external fields, which have
no analogue in \cite{Wulkenhaar:2001sq}. These graphs are computed to
\begin{fmffile}{fmfnonexp}
\begin{align}
\Gamma^{(1,1)}_{\bar{\rho}c\psi}(q,p,r) 
&= 
\parbox{40mm}{\begin{picture}(40,20)
\put(0,0){\begin{fmfgraph}(40,20)
\fmfleft{l1,l2}
\fmftop{t}
\fmfright{r}
\fmf{fermion}{r,i1,i2}
\fmf{dots}{i2,l2}
\fmf{phantom_arrow,tension=0}{i2,l2}
\fmf{scalar}{l1,i2}
\fmffreeze
\fmf{photon,left=0.5}{t,i1}
\fmf{phantom_arrow,left=0.5}{t,i1}
\fmf{photon,right=0.5}{t,i2}
\fmf{phantom_arrow,right=0.5}{t,i2}
\fmfdot{i2}
\end{fmfgraph}}
\put(8,2){\mbox{\small$p$}}
\put(0,15){\mbox{\small$-q$}}
\put(35,6){\mbox{\small$r$}}
\put(25,19){\mbox{\small$k$}}
\put(9,19){\mbox{\small$-k$}}
\put(16,6){\mbox{\small$k{+}r$}}
\end{picture}} 
\quad
= 
\frac{\hbar g^2}{(4\pi)^2\varepsilon} \theta^{\alpha\beta} p_\alpha 
\Big( -\frac{1}{4} r_\beta + \frac{1}{4}\gamma_{\beta\nu}
r^\nu - \frac{1}{2} m \gamma_\beta \Big)\;,
\label{G11rcj}
\\
\Gamma^{(1,1)}_{\bar{\psi}c\rho}(q,p,r) &= 
\parbox{40mm}{\begin{picture}(40,20)
\put(0,0){\begin{fmfgraph}(40,20)
\fmfright{r1,r2}
\fmftop{t}
\fmfleft{l}
\fmf{fermion}{i2,i1,l}
\fmf{dots}{r1,i2}
\fmf{phantom_arrow,tension=0}{r1,i2}
\fmf{scalar}{r2,i2}
\fmffreeze
\fmf{photon,left=0.5}{t,i2}
\fmf{phantom_arrow,left=0.5}{t,i2}
\fmf{photon,right=0.5}{t,i1}
\fmf{phantom_arrow,right=0.5}{t,i1}
\fmfdot{i2}
\end{fmfgraph}}
\put(35,15){\mbox{\small$p$}}
\put(2,6){\mbox{\small$-q$}}
\put(35,4){\mbox{\small$r$}}
\put(26,19){\mbox{\small$k$}}
\put(10,19){\mbox{\small$-k$}}
\put(14,6){\mbox{\small$k{-}q$}}
\end{picture}}  
\quad
= 
\frac{\hbar g^2}{(4\pi)^2\varepsilon}  \theta^{\alpha\beta} p_\alpha 
\Big( -\frac{1}{4} q_\beta - \frac{1}{4}
\gamma_{\beta\nu} q^\nu + \frac{1}{2} m \gamma_\beta \Big)\,,
\label{G11jcr}
\\
\Gamma^{(1,1)\nu}_{A\bar{\rho}c\psi}(q,r,p,s) &= 
\parbox{40mm}{\begin{picture}(40,20)
\put(0,0){\begin{fmfgraph}(40,20)
\fmfleft{l1,l2}
\fmftop{t}
\fmfright{b,r,b1}
\fmf{fermion}{r,i1,i3,i2}
\fmf{dots}{i2,l2}
\fmf{phantom_arrow,tension=0}{i2,l2}
\fmf{scalar}{l1,i2}
\fmffreeze
\fmf{photon,left=0.5}{t,i1}
\fmf{phantom_arrow,left=0.5}{t,i1}
\fmf{photon,right=0.5}{t,i2}
\fmf{phantom_arrow,right=0.5}{t,i2}
\fmf{photon}{b,i3}
\fmf{phantom_arrow}{b,i3}
\fmfdot{i2}
\end{fmfgraph}}
\put(8,2){\mbox{\small$p$}}
\put(0,15){\mbox{\small$-r$}}
\put(35,6){\mbox{\small$s$}}
\put(25,0){\mbox{\small$q,\nu$}}
\put(27,19){\mbox{\small$k{-}s$}}
\put(6,20){\mbox{\small$s{-}k$}}
\put(13,6){\mbox{\small$k{+}q$}}
\put(25,11.5){\mbox{\small$k$}}
\end{picture}}  
\quad
=
\frac{\hbar g^2}{(4\pi)^2\varepsilon} \theta^{\alpha\beta} p_\alpha 
\Big( -\frac{1}{4} \delta_\beta^\nu + \frac{1}{4} \gamma_\beta^{~\nu} 
\Big)\;,
\label{G11Arcj}
\\[3ex]
\Gamma^{(1,1)\nu}_{A\bar{\psi}c\rho}(q,r,p,s) &= 
\parbox{40mm}{\begin{picture}(40,20)
\put(0,0){\begin{fmfgraph}(40,20)
\fmfright{r1,r2}
\fmftop{t}
\fmfleft{b,l,b1}
\fmf{fermion}{i2,i3,i1,l}
\fmf{dots}{r1,i2}
\fmf{phantom_arrow,tension=0}{r1,i2}
\fmf{scalar}{r2,i2}
\fmffreeze
\fmf{photon,left=0.5}{t,i2}
\fmf{phantom_arrow,left=0.5}{t,i2}
\fmf{photon,right=0.5}{t,i1}
\fmf{phantom_arrow,right=0.5}{t,i1}
\fmf{photon}{b,i3}
\fmf{phantom_arrow}{b,i3}
\fmfdot{i2}
\end{fmfgraph}}
\put(2,6){\mbox{\small$-r$}}
\put(35,6){\mbox{\small$s$}}
\put(36,15){\mbox{\small$p$}}
\put(12,1){\mbox{\small$q,\nu$}}
\put(26,20){\mbox{\small$k{+}r$}}
\put(3,19){\mbox{\small$-r{-}k$}}
\put(20,6){\mbox{\small$k{-}q$}}
\put(13,11.5){\mbox{\small$k$}}
\end{picture}} 
\quad
=
\frac{\hbar g^2}{(4\pi)^2\varepsilon} \theta^{\alpha\beta} p_\alpha 
\Big( \frac{1}{4} \delta_\beta^\nu + \frac{1}{4} \gamma_\beta^{~\nu} 
\Big)\;.
\label{G11Ajcr}
\end{align}
The external fields $\bar{\rho},\rho$ are symbolised by dotted lines
and the ghost $c$ by dashed lines, everything else is as in
\cite{Wulkenhaar:2001sq}. A vertex with a dot is of first order in
$\theta$. 

First, the $(n{=}1,\ell{=}1)$ Slavnov-Taylor identities (\ref{STAjj})
and (\ref{STAA})--(\ref{STArcj}) are fulfilled, as already expected
from general considerations. For us the importance of these identities
consists in testing the graph computations performed by a
$\mathit{Mathematica}^{TM}$ program \cite{Ertl}. Next, the one-loop
divergent Green's functions at first order in $\theta$ are
considerably \emph{different} from their tree-level form. The question
is then how many of these divergences can be removed by a field
redefinition.

\subsection{Field redefinitions}

A field redefinition $\mathcal{F}$ must preserve the
Slavnov-Taylor identity, hence we have to require 
\begin{align}
  \mathcal{S} (\mathcal{F} \Gamma)&=0\;, &
\mathcal{F} &= \sum_{i} \int \Psi_i[\Phi_j] \frac{\delta}{\delta \Phi_i}\;,
\end{align}
where the functional $\Psi_i[\Phi_j]$ of the fields $\Phi_j$ must be
of the same power-counting dimension, ghost charge and hermiticity as
the field $\Phi_i$. We make the ansatz 
\begin{align}
\mathcal{F} \psi &= \psi -\frac{1}{2} \tau \theta^{\alpha\beta} A_\alpha
\partial_\beta \psi + \frac{\mathrm{i}}{4} \tau \theta^{\alpha\beta}
m A_\alpha \gamma_\beta \psi 
+ \frac{3}{8} \tau' \theta^{\alpha\beta} F^{\mu\nu}
\gamma_{\mu\nu\alpha\beta} \psi \;,
\label{Fj}
\\
\mathcal{F} \bar{\psi} &= \bar{\psi} -\frac{1}{2} \tau 
\theta^{\alpha\beta} \partial_\beta \bar{\psi} A_\alpha
- \frac{\mathrm{i}}{4} \tau \theta^{\alpha\beta}
\bar{\psi} \gamma_\beta m A_\alpha 
+ \frac{3}{8} \tau' \theta^{\alpha\beta} \bar{\psi} F^{\mu\nu}
\gamma_{\mu\nu\alpha\beta} \;,
\\
\mathcal{F} \rho &= \rho -\frac{1}{2} \tau \theta^{\alpha\beta} 
\partial_\beta (A_\alpha \rho)  
+ \frac{\mathrm{i}}{4} \tau \theta^{\alpha\beta}
m A_\alpha \gamma_\beta \rho 
- \frac{3}{8} \tau' \theta^{\alpha\beta} F^{\mu\nu}
\gamma_{\mu\nu\alpha\beta} \rho \;,
\\
\mathcal{F} \bar{\rho} &= \bar{\rho} -\frac{1}{2} \tau 
\theta^{\alpha\beta} \partial_\beta (\bar{\rho} A_\alpha)
- \frac{\mathrm{i}}{4} \tau \theta^{\alpha\beta}
\bar{\rho} \gamma_\beta m A_\alpha 
- \frac{3}{8} \tau' \theta^{\alpha\beta} \bar{\rho} F^{\mu\nu}
\gamma_{\mu\nu\alpha\beta} \;,
\\
\mathcal{F} \sigma^\mu &=  \sigma^\mu + \theta^{\mu\beta} 
\bar{\rho} \Big(\frac{1}{4} \tau (\delta^\nu_\beta -
\gamma_\beta^{~\nu})(\partial_\nu \psi -\mathrm{i} A_\nu \psi) 
+ \frac{\mathrm{i}}{2} \tau A_\beta \psi 
- \frac{\mathrm{i}}{2} \tau \gamma_\beta m \psi\Big)
\nonumber
\\
& \qquad - \theta^{\mu\beta} 
\Big(\frac{1}{4}  \tau (\partial_\nu \bar{\psi} 
+\mathrm{i} \bar{\psi} A_\nu) 
(\delta^\nu_\beta + \gamma_\beta^{~\nu})
- \frac{\mathrm{i}}{2} \tau \bar{\psi} A_\beta 
+ \frac{\mathrm{i}}{2}  \tau \bar{\psi} \gamma_\beta m \Big) \rho\;,
\\
\mathcal{F} A_\mu &= A_\mu\;,\qquad 
\mathcal{F} c = c\;,\qquad 
\mathcal{F} \kappa = \kappa\;,\qquad 
\mathcal{F} \bar{c} = \bar{c}\;,\qquad 
\mathcal{F} B = B\;,
\label{FB}
\end{align}
which leads to 
\begin{align}
\mathcal{F} (\Gamma^{(0,0)}) = \Gamma^{(0,0)}  
& + \tau \theta^{\alpha\beta} \Big(
-\frac{1}{2} \bar{\psi} \mathrm{i}\gamma^\mu 
\partial_\mu A_\alpha \partial_\beta \psi 
+ \frac{1}{2} \bar{\psi}\gamma^\mu A_\alpha 
\partial_\beta A_\mu \psi
-\frac{1}{4} 
\bar{\psi}\mathrm{i}\gamma^\mu F_{\alpha\beta} D_\mu \psi 
\nonumber
\\*
& 
\qquad +\frac{3}{8} \bar{\psi}m F_{\alpha\beta} \psi 
+ \frac{1}{4} 
\bar{\psi} m \gamma_\beta^{~\mu} (2 A_\alpha D_\mu \psi 
+ \partial_\mu A_\alpha \psi )\Big)
\nonumber
\\
& + \frac{3}{4} \tau' \theta^{\alpha\beta} \Big(
- \mathrm{i} \bar{\psi} \gamma_{\mu\nu\alpha} \partial_\beta
F^{\mu\nu} \psi 
+ \mathrm{i} \bar{\psi} \gamma_{\mu\alpha\beta} \partial_\nu
F^{\nu\mu} \psi 
- \bar{\psi} m \gamma_{\mu\nu\alpha\beta} F^{\mu\nu} \psi \Big)
\nonumber
\\*
& 
+ \tau \theta^{\alpha\beta} 
\bar{\rho} \partial_\alpha c \Big(\frac{1}{4} (\delta^\nu_\beta -
\gamma_\beta^{~\nu})(\partial_\nu \psi -\mathrm{i} A_\nu \psi) 
- \frac{\mathrm{i}}{2} \gamma_\beta m \psi\Big)
\nonumber
\\*
& + \tau \theta^{\alpha\beta} 
\Big(\frac{1}{4}  (\partial_\nu \bar{\psi} 
+\mathrm{i} \bar{\psi} A_\nu) 
(\delta^\nu_\beta + \gamma_\beta^{~\nu})
+ \frac{\mathrm{i}}{2} \bar{\psi} \gamma_\beta m \Big) 
\partial_\alpha c \rho  + \mathcal{O}(\theta^2)\;.
\end{align}
The corresponding Green's functions are
\begin{align}
(\mathcal{F}\Gamma)^{(1,0)\mu}_{A\bar{\psi}\psi}(p,q,r) 
&= \mathrm{i} \tau \theta^{\mu\beta} \Big(
-\frac{1}{2} (r_\nu p_\beta-p_\nu r_\beta) \gamma^\nu 
+ \frac{3}{4} m p_\beta 
- \frac{1}{4} m (2 r^\nu{+}p^\nu) \gamma_{\beta\nu}\Big) 
\nonumber
\\
& + \frac{3}{4} \mathrm{i} \tau' \theta^{\alpha\beta} \Big(
- 2 p_\beta p_\nu \gamma^{\mu\nu}_{~~\alpha}
- \gamma_{\nu\alpha\beta} (p^2 g^{\mu\nu} - p^\mu p^\nu)
- 2 m p_\nu \gamma^{\mu\nu}_{~~\alpha\beta}\Big) \;,
\\
(\mathcal{F}\Gamma)^{(1,0)\mu\nu}_{AA\bar{\psi}\psi}(p,q,r,s) &=
\mathrm{i} \tau \theta^{\alpha\beta}\Big( 
{-}\frac{1}{2} (p_\beta{+}q_\beta) (\delta^\mu_\alpha \gamma^\nu 
{+} \delta^\nu_\alpha \gamma^\mu) 
- \frac{1}{2} m (\delta^\mu_\alpha \gamma_\beta^{~\nu}
{+} \delta^\mu_\alpha \gamma_\beta^{~\mu})\Big)\;,
\\
(\mathcal{F}\Gamma)^{(1,0)}_{\bar{\rho}c\psi}(q,p,r) &=
\tau \theta^{\alpha\beta} p_\alpha \Big(
-\frac{1}{4} (r_\beta-\gamma_\beta^{~\nu} r_\nu) 
- \frac{1}{2} m \gamma_\beta\Big)\;,
\\
(\mathcal{F}\Gamma)^{(1,0)}_{\bar{\psi}c\rho}(q,p,r) &=
\tau \theta^{\alpha\beta} p_\alpha \Big(
-\frac{1}{4} (q_\beta+\gamma_\beta^{~\nu} q_\nu) 
+ \frac{1}{2} m \gamma_\beta\Big)\;,
\\
(\mathcal{F}\Gamma)^{(1,0)\nu}_{A\bar{\rho}c\psi}(q,r,p,s) &=
\tau \theta^{\alpha\beta} p_\alpha \Big(
-\frac{1}{4} (\delta_\beta^\nu -\gamma_\beta^{~\nu} ) 
\Big)\;,
\\
(\mathcal{F}\Gamma)^{(1,0)\nu}_{A\bar{\psi}c\rho}(q,r,p,s) &=
\tau \theta^{\alpha\beta} p_\alpha \Big(
\frac{1}{4} (\delta_\beta^\nu +\gamma_\beta^{~\nu} ) 
\Big)\;.
\end{align}
The Slavnov-Taylor identities (\ref{STAjj})--(\ref{STArcj}) are
verified. Now (\ref{G11Ajj}) and (\ref{G11rcj})--(\ref{G11Ajcr}) can
be rewritten as
\begin{align}
\Gamma^{(1,1)\mu}_{A\bar{\psi}\psi}(p,q,r)  
 &= \frac{\hbar  g^2}{(4\pi)^2 \varepsilon} 
\Big(\Big( \frac{1}{2} N_\psi + 0 N_A 
\Big)\Gamma^{(1,0)\mu}_{A\bar{\psi}\psi}(p,q,r)
+ \Big(\frac{\partial}{\partial\tau} 
+ \frac{\partial}{\partial\tau'} \Big) 
(\mathcal{F}\Gamma)^{(1,0)\mu}_{A\bar{\psi}\psi}(p,q,r)
\nonumber
\\*
& 
+ \mathrm{i}\theta^{\alpha\beta} \Big(
3 m \delta^\mu_\alpha p_\beta 
+ \frac{3}{2} m p_\nu \gamma^{\mu\nu}_{~~\alpha\beta} \Big)\Big)\;,
\label{G11AjjF}
\\
\Gamma^{(1,1)}_{\textit{ext.field}} &= 
\frac{\hbar g^2}{(4\pi)^2 \varepsilon} 
\frac{\partial}{\partial\tau} 
(\mathcal{F}\Gamma)^{(1,0)}_{\textit{ext.field}}\;,
\label{G11extF}
\end{align}
where ${}_{\textit{ext.field}}$ stands for
${}_{\bar{\rho}c\psi}(q,p,r)$, ${}_{\bar{\psi}c\rho}(q,p,r)$,
${}^\nu_{A\bar{\rho}c\psi}(q,r,p,s)$ and
${}^\nu_{A\bar{\psi}c\rho}(q,r,p,s)$. In other words, the one-loop
divergences in the Green's functions involving external fields and,
for $m=0$, in $\Gamma^{(1,1)\mu}_{A\bar{\psi}\psi}(p,q,r)$ can be
removed by field redefinitions. Due to the Slavnov-Taylor identity
these field redefinitions remove all one-loop divergences in
$\Gamma^{(1,1)\mu\nu}_{AA\bar{\psi}\psi}(p,q,r,s)$ and
$\Gamma^{(1,1)\mu\nu\rho}_{AAA\bar{\psi}\psi}(p,q,r,s,t)$ as well, and
$\Gamma^{(1,1)\mu_1\dots \mu_N}_{A\dots A}(p_1,\dots p_N)$ is
convergent for $N\in\{4,5,6\}$\footnote{\label{AA}Since all
  divergences in Green's functions involving external fields are
  removed by a field redefinition, see (\ref{G11extF}), the
  $(n{=}1,\ell{=}1)$ Slavnov-Taylor identity (\ref{STAAjj}) implies
  that the divergent part of
  $\Gamma^{(1,1)\mu\nu}_{AA\bar{\psi}\psi}(p,q,r,s)$ is transversal
  (contraction with $p_\mu$ yields zero) after the field redefinitions
  (\ref{Fj})--(\ref{FB}), because the remaining divergences in
  $\Gamma^{(1,1)\mu}_{A\bar{\psi}\psi}(p,q,r)$ are independent of $r$.
  Since $\Gamma^{(1,1)\mu\nu}_{AA\bar{\psi}\psi}(p,q,r,s)$ is linear
  in momentum variables and symmetric under $(p,\mu)\leftrightarrow
  (q,\nu)$, it must be zero. In the same way one proves
  $\Gamma^{(1,1)\mu\nu\rho}_{AAA\bar{\psi}\psi}(p,q,r,s,t)=0$ after
  the field redefinitions (\ref{Fj})--(\ref{FB}). Similarly, the
  photon $N$-point functions $\Gamma^{(1,1)\mu_1\dots \mu_N}_{A\dots
    A}(p_1,\dots p_N)$ for $N\in\{4,5,6\}$ are transversal in all
  momenta, but because they are at most quadratic (for $N=4$) in the
  momentum variables, they must vanish. This short proof shows that
  the computation of (\ref{G11AAjj}), (\ref{G11AAAjj}) and
  (\ref{G11AN}) was not necessary.}.  There remain only the divergence
in the electron four-point function (\ref{G11jjjj}) and the two mass
terms in (\ref{G11AjjF}). It is remarkable that these remaining
divergences coincide exactly (with the same numerical coefficients!)
with the result obtained in \cite{Wulkenhaar:2001sq} where the
electrons are Seiberg-Witten expanded! Moreover, there are no
divergences in the photon $N$-point functions
$\Gamma^{(1,1)\mu_1,\dots \mu_N}_{A\dots A}(p_1,\dots,p_N)$ after
\emph{the same} renormalisation of the coupling constant as in QED,
see (\ref{G01AA}). Again, this coincides with the results found in
\cite{Wulkenhaar:2001sq} where the fermions are Seiberg-Witten
expanded as well. This is a remarkable result: The physical (i.e.\ 
modulo field redefinitions) one-loop divergences are insensitive for
the choice of non-commutative or Seiberg-Witten expanded electrons in
$\theta$-expanded non-commutative QED.

\section{Expanding the action. [Case II]}
\label{sec4}

In this section we complete the first order analysis of the
Seiberg-Witten map on quantum level by leaving it out completely: We
repeat the analysis of the previous section without applying the
Seiberg-Witten map to the bosonic sector. The result is a `tilted'
BRST-symmetry in both bosonic and fermionic sectors leading to a tower
of symmetries involving both bosonic and fermionic actions.

\subsection{Classical analysis}

The expansion of the action (\ref{fullaction}), including the ghost
sector, is now performed according to
\begin{align}
\big(f\star g\big)(x) &= f(x)g(x) + \frac{{\rm{i}}}{2} 
\theta^{\a\b}\partial_\a f(x)\partial_\b g(x) 
+ \mathcal{O}(\q^{2}) \;,\nonumber
\\
\hat{\Phi} &= \Phi\;, \qquad \forall\;\hat{\Phi}\in \{\hat{A}_\mu,\hat{\psi},
\Hat{\Bar{\psi}},\hat{c},\hat{\bar{c}},\hat{B},
\hat{\Bar{\rho}},\hat{\rho},\hat{\sigma}^\mu,\hat{\kappa} \}\;.
\label{ExP}
\end{align}
This leads to the expanded action
\begin{align}
\Sigma^{\{n\}}_{\theta\textit{-exp}} 
&= \sum_{i=0}^n\Sigma^{(i)} \;,
\end{align}
which up to first order in $\theta$ (in which we are only interested for
now) reads
\begin{align}
\Sigma^{(0)}_{cl} &= \int d^4x \left(-\frac{1}{4g^2}
F_{\mu\nu} F^{\mu\nu} 
+\bar{\psi} \left( \mathrm{i} \gamma^\mu D_\mu -m \right) \psi\right)
\;,
\\
\Sigma^{(1)}_{cl}&= \int d^4x \left( -\frac{1}{2g^2}\theta^{\alpha\beta}
F_{\mu\nu}\partial_\alpha A^\mu \partial_\beta A^\nu
+\frac{\mathrm{i}}{2} \theta^{\alpha\beta} 
\bar{\psi} \gamma^\mu \partial_\alpha A_\mu \partial_\beta \psi
\right)\,,
\\
\Sigma^{(0)}_{gf} &=\int d^4x\;\Big( B \partial^\mu A_\mu 
- \bar{c}\partial^\mu\partial_\mu c +\frac{\a}{2}BB\Big)\;,
\label{sgf}
\\*
\Sigma^{(1)}_{gf} &=\int d^4x\;\Big(\theta^{\alpha\beta} 
\partial^\mu \bar{c}\, \partial_\alpha A_\mu\, \partial_\beta c  \Big)\;.
\label{inv}
\end{align}
We expand (\ref{BRST}) using (\ref{ExP}) to first order in $\theta$:
\begin{align}
s^{(0)}A_\m &= \partial_\m c \;,&
s^{(0)}c &= 0\;, 
\nonumber
\\*
s^{(0)}\j &= {\rm{i}}c\j \;,&
s^{(0)}\bar{\j}&= -{\rm{i}}\bar{\j} c \;,
\nonumber
\\
s^{(1)}A_\m &= \q^{\a\b}\partial_\a A_\m\partial_\b c  \;, &
s^{(1)}c &= -\frac{1}{2} \q^{\a\b}(\partial_\a c )(\partial_\b c) \;, 
\nonumber
\\*
s^{(1)}\j &=  - \frac{1}{2}\q^{\a\b}\partial_\a c\partial_\b\j \;, &
s^{(1)}\bar{\j} &= -\frac{1}{2}\q^{\a\b}
\partial_\b\bar{\j}\partial_\a c\;. 
\label{bbb}
\end{align}
The above transformations are $\theta$-graded in both bosonic and
fermionic sectors. The $\theta$-expanded BRST-transformations
(\ref{bbb}) fulfil (\ref{nr1}) and (\ref{nr2}).  Again we also expand
the term with external fields leading to (\ref{EXTER}) with the
BRST-transformations defined in (\ref{bbb}).  The full tree-level
generating functional is defined by (\ref{XX}), now with the classical
and gauge-fixing actions given by (\ref{inv}). 

Again the full set of BRST symmetries must be expressed by 
Slavnov-Taylor identities (\ref{ST}) and (\ref{STAjj})--(\ref{STArcj}). 

\subsection{The tree-level Green's functions}

At order $n{=}0$ in $\theta$ the tree-level Green's functions of [Case
II] are clearly the same as before (\ref{G00sc}). At order $n{=}1$ in
$\theta$ we now have the following non-vanishing tree-level Green's
functions:
\begin{align}
\Gamma^{(1,0)\mu}_{A\bar{\psi}\psi}(p,q,r,s) &= -\frac{\mathrm{i}}{2} 
\theta^{\alpha\beta} p_\alpha r_\beta \gamma^\mu\;,
\\
\Gamma^{(1,0)\mu}_{\bar{c}Ac}(p,q,r) &= \mathrm{i} 
\theta^{\alpha\beta} p^\mu q_\alpha r_\beta\;,
\\
\Gamma^{(1,0)\mu\nu\rho}_{AAA}(p,q,r) &= \frac{\mathrm{i}}{g^2}
\theta^{\alpha\beta} \Big( 
 (g^{\mu\nu} p^\rho - g^{\rho\mu} p^\nu) q_\alpha r_\beta 
+(g^{\nu\rho} q^\mu - g^{\mu\nu} q^\rho) r_\alpha p_\beta 
\nonumber
\\*
& \qquad\quad 
+(g^{\rho\mu} r^\nu - g^{\nu\rho} r^\mu) p_\alpha q_\beta \Big)\;,
\\
\Gamma^{(1,0)\nu}_{\mu;A\sigma c}(q,r,p) &= \delta_\mu^\nu 
\theta^{\alpha\beta} p_\alpha q_\beta\;,
\\
\Gamma^{(1,0)}_{\bar{\rho}c \psi}(q,p,r) &= \frac{1}{2} 
\theta^{\alpha\beta} p_\alpha r_\beta\;,
\\
\Gamma^{(1,0)}_{\bar{\psi}c \rho}(q,p,r) &= \frac{1}{2} 
\theta^{\alpha\beta} p_\alpha q_\beta\;,
\\
\Gamma^{(1,0)}_{\kappa cc}(p,q,r) &= 
\theta^{\alpha\beta} q_\alpha r_\beta\;.
\end{align}
It is straightforward to check the $(n{=}1,\ell{=}0)$ Slavnov-Taylor
identities (\ref{STAjj})--(\ref{STArcj}).

\subsection{One-loop computation}

The one-loop results for order $n{=}0$ in $\theta$ are the same as
before (\ref{G01jj})--(\ref{G01AA}). At order $n{=}1$ in $\theta$ we
find the following divergent Green's functions in analytic
regularisation (using again the Feynman gauge $\alpha=g^2$): 
\begin{align}
\Gamma^{(1,1)}_{\bar{\psi}\psi}(q,p) &= 0\;,
\label{g11jj}
\\*
\Gamma^{(1,1)\mu}_{A\bar{\psi}\psi}(p,q,r) &= \frac{\hbar
  g^2}{(4\pi)^2 \varepsilon} \Big(
\Big(\frac{1}{2} N_\psi + 0 N_A\Big)
\Gamma^{(1,0)\mu}_{A\bar{\psi}\psi}(p,q,r) 
\nonumber
\\*
& \qquad +\mathrm{i}\theta^{\alpha\beta} 
\Big(
- \frac{1}{2} p_\alpha r_\beta \gamma^\mu 
- \frac{1}{4} (2 r^\mu + p^\mu) p_\beta \gamma_\alpha
- \frac{1}{4} \delta^\mu_\alpha (2 r_\nu + p_\nu) p_\beta \gamma^\nu 
\nonumber
\\*
&
\qquad\qquad 
-\frac{5}{4} p_\nu p_\beta \gamma^{\mu\nu}_{~~\alpha}
+ \frac{7}{2} m \delta^\mu_\alpha p_\beta \Big)\Big)\;,
\label{g11Ajj}
\\
\Gamma^{(1,1)\mu\nu}_{AA\bar{\psi}\psi}(p,q,r,s) &= \dots\;,
\label{g11AAjj}
\\
\Gamma^{(1,1)\mu\nu\rho}_{AAA\bar{\psi}\psi}(p,q,r,s,t) &= \dots\;,
\label{g11AAAjj}
\\
\Gamma^{(1,1)}_{\bar{\psi}\psi;\bar{\psi}\psi}(p,q,r,s) &= 0\;,
\label{g11jjjj}
\\
\Gamma^{(1,1)\mu\nu}_{AA}(p,q) &= 0\;,
\label{g11AA}
\\
\Gamma^{(1,1)\mu\nu\rho}_{AAA}(p,q,r) &= \frac{\hbar
  g^2}{(4\pi)^2\varepsilon} \Big( -\frac{4}{3} g^2
\frac{\partial}{\partial g^2} + 0 N_A \Big) 
\Gamma^{(1,0)\mu\nu\rho}_{AAA}(p,q,r)\;,
\label{g11AAA}
\\
\Gamma^{(1,1)\mu_1\dots \mu_N}_{A\dots A}(p_1,\dots & p_N) = \dots \;,
\qquad N\in\{4,5,6\}\;,
\label{g11AN}
\\
\Gamma^{(1,1)}_{\bar{\rho}c\psi}(q,p,r) &= 
\parbox{40mm}{\begin{picture}(40,30)
\put(0,0){\begin{fmfgraph}(40,30)
\fmfright{r1,r2}
\fmfleft{l}
\fmf{dots,tension=2}{i2,r2}
\fmf{phantom_arrow,tension=1}{i2,r2}
\fmf{scalar,tension=3}{r1,i1}
\fmf{fermion,tension=3}{l,i3}
\fmf{fermion}{i3,i2}
\fmf{scalar}{i1,i2}
\fmf{phantom}{i1,i3}
\fmffreeze
\fmf{photon}{i4,i1}
\fmf{phantom_arrow}{i4,i1}
\fmf{photon}{i4,i3}
\fmf{phantom_arrow}{i4,i3}
\fmfdot{i1}
\end{fmfgraph}}
\put(2,11){\mbox{$r$}}
\put(9,9){\mbox{$k,\sigma$}}
\put(12,20){\mbox{$k{+}r$}}
\put(26,26){\mbox{$-q$}}
\put(18,5){\mbox{${-}k,\rho$}}
\put(29,1){\mbox{$p$}}
\put(32,15){\mbox{$p{-}k$}}
\end{picture}}
\quad
=
\frac{\hbar g^2}{(4\pi)^2\varepsilon}\theta^{\alpha\beta} p_\alpha 
\Big( -\frac{1}{4} q_\beta - \frac{1}{4} m \gamma_\beta 
- \frac{1}{4} \gamma_{\beta\nu} q^\nu\Big)\;,
\label{g11rcj}
\\
\Gamma^{(1,1)}_{\bar{\psi}c\rho}(q,p,r) &= 
\parbox{40mm}{\begin{picture}(40,30)
\put(0,0){\begin{fmfgraph}(40,30)
\fmfright{r1,r2}
\fmfleft{l}
\fmf{dots,tension=2}{l,i3}
\fmf{phantom_arrow,tension=1}{l,i3}
\fmf{scalar,tension=3}{r1,i1}
\fmf{fermion,tension=3}{i2,r2}
\fmf{fermion}{i3,i2}
\fmf{scalar}{i1,i3}
\fmf{phantom}{i1,i2}
\fmffreeze
\fmf{photon}{i4,i1}
\fmf{phantom_arrow}{i4,i1}
\fmf{photon}{i4,i2}
\fmf{phantom_arrow}{i4,i2}
\fmfdot{i1}
\end{fmfgraph}}
\put(2,11){\mbox{$r$}}
\put(12,20){\mbox{$k{-}q$}}
\put(26,26){\mbox{$-q$}}
\put(17,7){\mbox{$k{+}p$}}
\put(29,1){\mbox{$p$}}
\put(32,18){\mbox{${-}k,\rho$}}
\put(32,11){\mbox{$k,\sigma$}}
\end{picture}}
\quad
=
\frac{\hbar  g^2}{(4\pi)^2\varepsilon}\theta^{\alpha\beta} p_\alpha 
\Big( - \frac{1}{4} r_\beta + \frac{1}{4} m \gamma_\beta 
+ \frac{1}{4} \gamma_{\beta\nu} r^\nu\Big)\;,
\label{g11jcr}
\\
\Gamma^{(1,1)\nu}_{A\bar{\rho}c\psi}(q,r,p,s) &= 
\parbox{40mm}{\begin{picture}(40,30)
\put(0,0){\begin{fmfgraph}(40,30)
\fmfright{r1,r2}
\fmfleft{l1,l2}
\fmf{dots,tension=2}{i2,r2}
\fmf{phantom_arrow,tension=1}{i2,r2}
\fmf{scalar,tension=3}{r1,i1}
\fmf{fermion,tension=3}{l1,i3}
\fmf{fermion}{i3,i5,i2}
\fmf{scalar}{i1,i2}
\fmf{phantom}{i1,i3}
\fmf{photon,tension=2}{l2,i5}
\fmf{phantom_arrow}{l2,i5}
\fmffreeze
\fmf{photon}{i4,i1}
\fmf{phantom_arrow}{i4,i1}
\fmf{photon}{i4,i3}
\fmf{phantom_arrow}{i4,i3}
\fmfdot{i1}
\end{fmfgraph}}
\put(3,3){\mbox{$s$}}
\put(1,13){\mbox{$k{+}s$}}
\put(9,27){\mbox{$q,\nu$}}
\put(11,2){\mbox{$k,\sigma$}}
\put(14,20){\mbox{$k{+}q{+}s$}}
\put(25,27){\mbox{$-r$}}
\put(19,2){\mbox{${-}k,\rho$}}
\put(33,5){\mbox{$p$}}
\put(32,15){\mbox{$p{-}k$}}
\end{picture}}
\quad
=\frac{\hbar
  g^2}{(4\pi)^2\varepsilon}\theta^{\alpha\beta} p_\alpha \Big(
\frac{1}{4} \delta_\beta^\nu
+ \frac{1}{4} \gamma_\beta^{~\nu}\Big)\;,
\label{g11Arcj}
\\[1ex]
\Gamma^{(1,1)\nu}_{A\bar{\psi}c\rho}(q,r,p,s) &= 
\parbox{40mm}{\begin{picture}(40,30)
\put(0,0){\begin{fmfgraph}(40,30)
\fmfright{r1,r2}
\fmfleft{l1,l2}
\fmf{dots,tension=2}{l1,i3}
\fmf{phantom_arrow,tension=1}{l1,i3}
\fmf{scalar,tension=3}{r1,i1}
\fmf{fermion,tension=3}{i2,r2}
\fmf{fermion}{i3,i5,i2}
\fmf{scalar}{i1,i3}
\fmf{phantom}{i1,i2}
\fmf{photon,tension=2}{l2,i5}
\fmf{phantom_arrow}{l2,i5}
\fmffreeze
\fmf{photon}{i4,i1}
\fmf{phantom_arrow}{i4,i1}
\fmf{photon}{i4,i2}
\fmf{phantom_arrow}{i4,i2}
\fmfdot{i1}
\end{fmfgraph}}
\put(3,3){\mbox{$s$}}
\put(9,27){\mbox{$q,\nu$}}
\put(26,27){\mbox{$-r$}}
\put(16,20){\mbox{$k{-}r$}}
\put(33,5){\mbox{$p$}}
\put(16,2){\mbox{$k{+}p$}}
\put(-3,18){\mbox{$k{+}p{+}s$}}
\put(31,18){\mbox{${-}k,\rho$}}
\put(32,11){\mbox{$k,\sigma$}}
\end{picture}}
\quad
=\frac{\hbar
  g^2}{(4\pi)^2\varepsilon}\theta^{\alpha\beta} p_\alpha \Big(
-\frac{1}{4} \delta_\beta^\nu 
+ \frac{1}{4} \gamma_\beta^{~\nu} \Big)\;.
\label{g11Ajcr}
\end{align}
The $(n{=}1,\ell{=}1)$ Slavnov-Taylor identities (\ref{STAjj}) and
(\ref{STAA})--(\ref{STArcj}) are verified.

\end{fmffile}

\subsection{Field redefinitions}

We try again to absorb the divergences in
(\ref{g11rcj})--(\ref{g11Ajcr}) and most of (\ref{g11Ajj}) by field
redefinitions. We make the ansatz
\begin{align}
\mathcal{F} \psi &= \psi + \theta^{\alpha\beta} \Big(
-\frac{1}{4} \tau \gamma_\alpha^{~\nu} \partial_\beta A_\nu \psi 
+ \frac{3}{8} \tau' \gamma_{\mu\nu\alpha\beta} F^{\mu\nu} \psi
-\frac{1}{8} \tau'' F_{\alpha\beta} \psi
\Big)\;,
\label{fj}
\\
\mathcal{F} \bar{\psi} &= \bar{\psi} + \theta^{\alpha\beta} \Big(
\frac{1}{4} \tau \bar{\psi} \gamma_\alpha^{~\nu}  \partial_\beta A_\nu
+ \frac{3}{8} \tau' \bar{\psi} \gamma_{\mu\nu\alpha\beta} F^{\mu\nu} 
-\frac{1}{8} \tau'' \bar{\psi} F_{\alpha\beta} 
\Big)\;,
\\
\mathcal{F} \bar{\rho} &= \bar{\rho} + \theta^{\alpha\beta} \Big(
\frac{1}{4} \tau \bar{\rho} \gamma_\alpha^{~\nu} \partial_\beta A_\nu 
- \frac{3}{8} \bar{\rho} \tau' \gamma_{\mu\nu\alpha\beta} F^{\mu\nu} 
+\frac{1}{8} \tau'' \bar{\rho} F_{\alpha\beta} 
\Big)\;,
\\
\mathcal{F} \rho &= \rho + \theta^{\alpha\beta} \Big(
-\frac{1}{4} \tau \gamma_\alpha^{~\nu}  \partial_\beta A_\nu \rho 
- \frac{3}{8} \tau' \gamma_{\mu\nu\alpha\beta} F^{\mu\nu} \rho
+\frac{1}{8} \tau'' F_{\alpha\beta} \rho
\Big)\;,
\\
\mathcal{F} A_\mu &= A_\mu 
-\frac{3}{4} \mathrm{i} g^2 \tau''' \theta^{\alpha\beta} 
\bar{\psi} \gamma_{\mu\alpha\beta} \psi\;,
\\
\mathcal{F} \sigma^\mu &= \sigma^\mu 
+ \frac{1}{4} \tau \theta^{\mu\beta} \Big(
(\partial_\nu \bar{\rho}+\mathrm{i} \bar{\rho} A_\nu) 
(\delta^\nu_\beta+\gamma_\beta^{~\nu})\psi  
- \mathrm{i} \bar{\rho} m \gamma_\beta \psi
\nonumber
\\*
&\qquad \quad - \bar{\psi} (\delta^\nu_\beta-\gamma_\beta^{~\nu}) 
(\partial_\nu \rho -\mathrm{i} A_\nu \rho) 
- \mathrm{i} \bar{\psi} m \gamma_\beta \rho\Big)\;,
\\
\mathcal{F} c &= c\;,\qquad 
\mathcal{F} \kappa = \kappa\;,\qquad 
\mathcal{F} \bar{c} = \bar{c}\;,\qquad 
\mathcal{F} B = B\;,
\end{align}
which gives 
\begin{align}
\mathcal{F} (\Gamma^{(0,0)}) = \Gamma^{(0,0)}  
& + \tau \theta^{\alpha\beta} \Big(
- \frac{1}{2} \bar{\psi} \mathrm{i}\gamma^\nu
\partial_\beta A_\nu \partial_\alpha \psi 
+ \frac{1}{4} \bar{\psi} \mathrm{i}\gamma_\alpha
\partial^\nu \partial_\beta A_\nu \psi 
+ \frac{1}{2} \bar{\psi} \mathrm{i}\gamma_\alpha
 \partial_\beta A_\nu \partial^\nu \psi 
\nonumber
\\
& 
\qquad + \frac{1}{4} \bar{\psi} \mathrm{i}\gamma^{\mu\nu}_{~~\alpha}
\partial_\mu \partial_\beta A_\nu \psi 
-\frac{1}{2} \bar{\psi} \gamma^\nu A_\alpha \partial_\beta A_\nu \psi
+\frac{1}{2} \bar{\psi} \gamma_\alpha A^\nu \partial_\beta A_\nu \psi
\Big)
\nonumber
\\
& + \frac{3}{4} \tau' \theta^{\alpha\beta} \Big(
- \mathrm{i} \bar{\psi} \gamma_{\mu\nu\alpha} \partial_\beta
F^{\mu\nu} \psi 
+ \mathrm{i} \bar{\psi} \gamma_{\mu\alpha\beta} \partial_\nu
F^{\nu\mu} \psi 
- \bar{\psi} m \gamma_{\mu\nu\alpha\beta} F^{\mu\nu} \psi \Big)
\nonumber
\\
& 
+ \tau'' \theta^{\alpha\beta} \Big(
-\frac{1}{8} \bar{\psi} \mathrm{i} \gamma^\mu 
(\partial_\mu F_{\alpha\beta}
\psi + 2 F_{\alpha\beta} D_\mu \psi)
+\frac{1}{4} \bar{\psi} m F_{\alpha\beta} \psi 
\Big)
\nonumber
\\
& - \frac{3}{4} \tau'''\theta^{\alpha\beta} \Big(
\bar{\psi} \mathrm{i} \gamma_{\mu\alpha\beta} 
(\partial_\nu F^{\nu\mu} -g^2 \partial^\mu B) \psi 
+ g^2 (\bar{\psi} \gamma^\mu \psi)(\bar{\psi}\mathrm{i} 
\gamma_{\mu\alpha\beta}\psi) \Big)
\nonumber
\\
&+ \frac{1}{4} \tau \theta^{\alpha\beta} \Big(
(\partial_\nu \bar{\rho}+\mathrm{i} \bar{\rho} A_\nu) 
(\delta^\nu_\beta+\gamma_\beta^{~\nu}) \partial_\alpha c \psi  
- \mathrm{i} \bar{\rho} m \gamma_\beta \partial_\alpha c \psi
\nonumber
\\
&\qquad + \bar{\psi} \partial_\alpha c 
(\delta^\nu_\beta-\gamma_\beta^{~\nu}) 
(\partial_\nu \rho -\mathrm{i} A_\nu \rho) 
+ \mathrm{i} \bar{\psi} m \partial_\alpha c \gamma_\beta \rho\Big)
+ \mathcal{O}(\theta^2)\;.
\label{fg}
\end{align}
The corresponding Green's functions are
\begin{align}
(\mathcal{F} \Gamma)^{(1,0)\mu}_{A\bar{\psi}\psi}(p,q,r) &= 
\mathrm{i} \theta^{\alpha\beta} \Big(
-\frac{1}{2} \tau p_\alpha r_\beta \gamma^\mu 
-\frac{1}{4} \tau (p^\mu{+}2r^\mu) p_\beta \gamma_\alpha   
- \frac{1}{4} \tau'' (p_\nu{+}2r_\nu) p_\beta 
\delta^\mu_\alpha \gamma^\nu
\nonumber
\\
&
\qquad + \Big(\frac{1}{4}\tau {-}\frac{3}{2} \tau'\Big) 
p_\nu p_\beta \gamma^{\mu\nu}_{~~\alpha} 
+\frac{3}{4} (\tau'''-\tau') (p^2 g^{\mu\nu} {-} p^\mu p^\nu) 
\gamma_{\nu\alpha\beta}
\nonumber
\\
&\qquad
+ \frac{1}{2} \tau'' m p_\beta \delta^\mu_\alpha
-\frac{3}{2} \tau' m p^\nu \gamma^{\mu\nu}_{~~\alpha\beta}\Big)\;,
\\
(\mathcal{F} \Gamma)^{(1,0)\mu\nu}_{AA\bar{\psi}\psi}(p,q,r,s) &= 
\mathrm{i} \theta^{\alpha\beta} \Big(
\frac{1}{2} (\tau q_\beta -\tau'' p_\beta) \delta^\mu_\alpha \gamma^\nu 
+\frac{1}{2} (\tau p_\beta -\tau'' q_\beta) \delta^\nu_\alpha \gamma^\mu
- \frac{1}{2} \tau (p_\beta{+}q_\beta) g^{\mu\nu} \gamma_\alpha
\Big)\;,
\\
(\mathcal{F} \Gamma)^{(1,0)}_{\bar{\psi}\psi;\bar{\psi}\psi}(p,q,r,s)
&= -\frac{3}{4} \tau'''\mathrm{i} \theta^{\alpha\beta} g^2 
\gamma^\mu \otimes \gamma_{\mu\alpha\beta}\;,
\\
(\mathcal{F} \Gamma)^{(1,0)}_{B\bar{\psi}\psi}(p,q,r)
&= -\frac{3}{4} \tau'''\mathrm{i} \theta^{\alpha\beta} g^2 
p^\mu \gamma_{\mu\alpha\beta}\;,
\\
(\mathcal{F} \Gamma)^{(1,0)}_{\bar{\rho}c\psi}(q,p,r) &= 
\frac{1}{4}\tau \theta^{\alpha\beta} p_\alpha 
\Big( q_\nu (-\delta^\nu_\beta - \gamma_\beta^{~\nu})
- m \gamma_\beta \Big)\;,
\\
(\mathcal{F} \Gamma)^{(1,0)}_{\bar{\psi}c\rho}(q,p,r) &= 
\frac{1}{4}\tau \theta^{\alpha\beta} p_\alpha 
\Big( r_\nu (-\delta^\nu_\beta + \gamma_\beta^{~\nu})
+ m \gamma_\beta \Big)\;,
\\
(\mathcal{F} \Gamma)^{(1,0)\nu}_{A\bar{\rho}c\psi}(q,r,p,s) &= 
\frac{1}{4}\tau \theta^{\alpha\beta} p_\alpha 
q_\nu (\delta^\nu_\beta + \gamma_\beta^{~\nu})\;,
\\
(\mathcal{F} \Gamma)^{(1,0)}_{A\bar{\psi}c\rho}(q,r,p,s) &= 
\frac{1}{4}\tau \theta^{\alpha\beta} p_\alpha 
q_\nu (-\delta^\nu_\beta + \gamma_\beta^{~\nu})\;.
\end{align}
The $(n{=}1,\ell{=}0)$ Slavnov-Taylor identities
(\ref{STAjj})--(\ref{STArcj}) are verified. Now
(\ref{g11Ajj}), (\ref{g11jjjj}) and (\ref{g11rcj})--(\ref{g11Ajcr})
can be rewritten as
\begin{align}
\Gamma^{(1,1)\mu}_{A\bar{\psi}\psi}(p,q,r)  
 &= \frac{\hbar  g^2}{(4\pi)^2 \varepsilon} 
\Big(\Big( \frac{1}{2} N_\psi + 0 N_A 
\Big)\Gamma^{(1,0)\mu}_{A\bar{\psi}\psi}(p,q,r)
\nonumber
\\*
& + \Big(\frac{\partial}{\partial\tau} 
+ \frac{\partial}{\partial\tau'} 
+ \frac{\partial}{\partial\tau''} + \frac{\partial}{\partial\tau'''} 
\Big) 
(\mathcal{F}\Gamma)^{(1,0)\mu}_{A\bar{\psi}\psi}(p,q,r)
\nonumber
\\*
& 
+ \mathrm{i}\theta^{\alpha\beta} \Big(
3 m \delta^\mu_\alpha p_\beta 
+ \frac{3}{2} m p_\nu \gamma^{\mu\nu}_{~~\alpha\beta} \Big)\Big)\;,
\label{g11AjjF}
\\
\Gamma^{(1,1)}_{\bar{\psi}\psi;\bar{\psi}\psi}(p,q,r,s)  
&= \frac{\hbar  g^2}{(4\pi)^2 \varepsilon} \Big(
\frac{\partial}{\partial \tau'''} 
(\mathcal{F}\Gamma)^{(1,0)}_{\bar{\psi}\psi;\bar{\psi}\psi}(p,q,r,s)
+ \frac{3}{4} \mathrm{i} \theta^{\alpha\beta} g^2 \gamma^\mu \otimes
\gamma_{\mu\alpha\beta} \Big)\;,
\label{g11jjjjF}
\\
\Gamma^{(1,1)}_{B\bar{\psi}\psi}(p,q,r)  
&= \frac{\hbar  g^2}{(4\pi)^2 \varepsilon} \Big(
\frac{\partial}{\partial \tau'''} 
(\mathcal{F}\Gamma)^{(1,0)}_{B\bar{\psi}\psi}(p,q,r)
+ \frac{3}{4} \mathrm{i} \theta^{\alpha\beta} g^2 p^\mu 
\gamma_{\mu\alpha\beta} \Big)\;,
\\
\Gamma^{(1,1)}_{\textit{ext.field}} &= 
\frac{\hbar g^2}{(4\pi)^2 \varepsilon} 
\frac{\partial}{\partial\tau} 
(\mathcal{F}\Gamma)^{(1,0)}_{\textit{ext.field}}\;,
\end{align}
where ${}_{\textit{ext.field}}$ stands for
${}_{\bar{\rho}c\psi}(q,p,r)$, ${}_{\bar{\psi}c\rho}(q,p,r)$,
${}^\nu_{A\bar{\rho}c\psi}(q,r,p,s)$ and
${}^\nu_{A\bar{\psi}c\rho}(q,r,p,s)$. Thus, the result after field
redefinitions is the same as in [Case I] and \cite{Wulkenhaar:2001sq},
provided that a $\hbar$-renormalisation of the tree-level gauge-fixing
action $\Sigma^{(0)}_{gf}$ from (\ref{sgf}) to 
\begin{align}
\Sigma^{\prime(0)}_{gf} &=\int d^4x\;\Big( 
B \partial^\mu \Big(A_\mu - \frac{3}{4} g^2 \frac{\hbar g^2}{(4\pi)^2
  \varepsilon} \theta^{\alpha\beta} \bar{\psi}
\gamma_{\mu\alpha\beta} \psi \Big)
- \bar{c} \partial^\mu \partial_\mu c  +\frac{\alpha}{2} BB\Big)
\end{align}
is performed. In summary, up to field redefinitions the one-loop
computations of Green's functions up to first order in $\theta$ are
completely independent of the application of the Seiberg-Witten map 

(1) to both electrons and photons \cite{Wulkenhaar:2001sq}, 

(2) to photons only [Case I], or 

(3) to neither photons nor electrons [Case II]. 
\\
In the next section we shall explain why this has to be the case.

First let us point out a possibility which we have overlooked in 
\cite{Wulkenhaar:2001sq} and which becomes apparent from the loop
calculation of [Case II]. Putting $\tau'=\tau'''=0$ in (\ref{fg}) we 
have instead of (\ref{g11AjjF}) and (\ref{g11jjjjF}) 
\begin{align}
\Gamma^{(1,1)\mu}_{A\bar{\psi}\psi}(p,q,r)  
 &= \frac{\hbar  g^2}{(4\pi)^2 \varepsilon} 
\Big(\Big( \frac{1}{2} N_\psi + 0 N_A 
\Big)\Gamma^{(1,0)\mu}_{A\bar{\psi}\psi}(p,q,r)
+ \Big(\frac{\partial}{\partial\tau} 
+ \frac{\partial}{\partial\tau''} \Big) 
(\mathcal{F}\Gamma)^{(1,0)\mu}_{A\bar{\psi}\psi}(p,q,r)
\nonumber
\\*
& 
+ \mathrm{i}\theta^{\alpha\beta} \Big(
3 m \delta^\mu_\alpha p_\beta 
- \frac{3}{2} p_\nu p_\beta \gamma^{\mu\nu}_{~~\alpha} \Big)\Big)\;,
\label{g11AjjF0}
\\
\Gamma^{(1,1)}_{\bar{\psi}\psi;\bar{\psi}\psi}(p,q,r,s)  
&= 0\;.
\label{g11jjjjF0}
\end{align}
The same result can obviously be achieved for the treatments of
\cite{Wulkenhaar:2001sq} and [Case I] as well. This is the minimal
field redefinition in the sense that only two non-absorbable one-loop
divergences remain. It is tempting to try an extended non-commutative
initial action
\begin{align}
\hat{\Sigma}^e_{\mathit{cl}} &=  \hat{\Sigma}_{\mathit{cl}} 
+ g_e \int \!d^4x\;  
\mathrm{i} \theta^{\alpha\beta}
\hat{\bar{\psi}} \star \gamma^{\mu\nu}_{~~\alpha} 
\hat{D}^{adj}_\beta \hat{F}_{\mu\nu}
\star \hat{\psi} \;,
\label{extended}
\\*
\hat{D}^{adj}_\beta \hat{F}_{\mu\nu} &=
\partial_\beta \hat{F}_{\mu\nu}
-\mathrm{i} [\hat{A}_\beta,\hat{F}_{\mu\nu}]_\star \;,
\nonumber
\end{align}
where $\hat{\Sigma}_{\mathit{cl}}$ was given in (\ref{theaction}) and
$g_e$ is a new coupling constant.
It turns out that all divergences generated by this extension term
are---apart from the trivial one due to the wave function
renormalisation of $\bar{\psi},\psi$---proportional to the electron
mass $m$.  In other words, in massless non-commutative QED the
$\theta$-expansion of (\ref{extended}) is one-loop renormalisable up
to first order in $\theta$ by the standard QED wave function and
electron charge renormalisations, the renormalisation 
\begin{equation}
g_e(\varepsilon) =  g_e + \frac{3}{4} \frac{\hbar g^2}{(4\pi)^2
  \varepsilon} 
\end{equation}
of the additional coupling constant $g_e$ and field redefinitions.

\section{General considerations: Change of variables}
\label{sec5}

In this section we further analyse NCYM theory expanded in $\q$. In
the following we shall leave the option open as to whether fermions
are included or not. Our starting point is a trivial expansion of
(\ref{theaction}) according to
\begin{align}
 (f\star g)(x) &= f(x)g(x) + \sum_{n=1}^\infty \frac{1}{n!} 
\Big(\frac{\mathrm{i}}{2}\Big)^n  
\theta^{\alpha_1\beta_1}\cdots \theta^{\alpha_n\beta_n}
\big(\partial_{\alpha_1}\dots \partial_{\alpha_n} f(x)\big) 
\big(\partial_{\beta_1}\dots \partial_{\beta_n} g(x)\big) \;,
\nonumber
\\
\hat{\Phi}_i &= \Phi_i'\;, \qquad \forall\;\hat{\Phi}_i\;,
\label{XX1}
\end{align}
where $\hat{\F}_i$ denotes {\it all}\/ fields of the theory, with the
index $i$ labelling spin and particle type.  
We reconsider the Seiberg-Witten map
\begin{align}
\Phi'_i &= \Phi_i + \Omega_i[\Phi]\;,
\label{redef}
\end{align}
where the field polynomial $\Omega_i[\Phi]$ is at least linear in
$\theta$, as a change of integration variables in the path integral
\begin{equation}
Z[J]=\mathcal{N} \int\big[\mathcal{D} \Phi'\big]
\;\mbox{exp} 
\Big(\frac{\mathrm{i}}{\hbar} \Gamma_{\mathit{cl}}[\Phi'] 
+ \frac{\mathrm{i}}{\hbar} J^i \Phi_i' \Big)\;.
\label{path}
\end{equation}
Here, $\mathcal{N}$ is a (ill-defined) normalisation factor and 
$\Gamma_{\mathit{cl}}[\Phi']$ is the gauge-fixed NCYM
action---possibly including fermions---expanded according to
(\ref{XX1}) in $\theta$. To improve the readability we omit space-time
integrals in $ J^i\Phi_i' \equiv \int d^4x \;J^i(x) \Phi_i'(x)$ as well
as in the sequel. We apply (\ref{redef}) to (\ref{path}) and
find
\begin{align}
Z[J]&= \mathcal{N} \int\big[\mathcal{D}\Phi\big] \; 
\det\left[\frac{\delta \Phi_j'}{\delta \Phi_k}\right]\;
\exp \Big( \frac{\mathrm{i}}{\hbar} \Gamma_{\mathit{cl}}[\Phi'[\Phi]]
+ \frac{\mathrm{i}}{\hbar} J^i \Phi_i'[\Phi] \Big)
\nonumber
\\*
&= \mathcal{N} \int\big[\mathcal{D} \Phi\big]
\big[\mathcal{D} \mathcal{C}\big]
\big[ \mathcal{D} \bar{\mathcal{C}}\big]\;
\exp \Big( \frac{\mathrm{i}}{\hbar} \Gamma_{\mathit{cl}}[\Phi'[\Phi]] 
+\bar{\mathcal{C}}^i \frac{\delta \Phi'_i}{
\delta \Phi_j} \mathcal{C}_j
\nonumber
\\*
&\hspace*{13em} 
+ \frac{\mathrm{i}}{\hbar} \Big(J^i \Phi_i 
+ J^i \Omega_i[\Phi] + \bar{\mathcal{C}}^i \mathcal{J}_i 
+ \bar{\mathcal{J}}^i \mathcal{C}_i \Big)\Big)
\bigg|_{\mathcal{J}=\bar{\mathcal{J}}=0}
\nonumber
\\
&\equiv \mathcal{N} \int\big[\mathcal{D} \Phi\big]
\big[\mathcal{D} \mathcal{C}\big]
\big[ \mathcal{D} \bar{\mathcal{C}}\big]\;
\exp \Big( \frac{\mathrm{i}}{\hbar} 
\tilde{\Gamma}_{\mathit{cl}}[\Phi,\mathcal{C},\bar{\mathcal{C}}] 
\nonumber
\\*
&\hspace*{13em} 
+ \frac{\mathrm{i}}{\hbar} \Big(J^i \Phi_i 
+ J^i \Omega_i[\Phi] + \bar{\mathcal{C}}^i \mathcal{J}_i 
+ \bar{\mathcal{J}}^i \mathcal{C}_i \Big)
\Big)\bigg|_{\mathcal{J}=\bar{\mathcal{J}}=0}\;.
\label{holl}
\end{align}
The ghosts and anti-ghosts $\mathcal{C}_i$ and $\bar{\mathcal{C}}_i$
are to be understood as `towers' of fields of mixed Grassmann grading
according to the actual field they couple to. The effect of the new ghost
sector introduced in (\ref{holl}) is of course to compensate for the
performed field redefinition in agreement with the equivalence theorem
\cite{Kamefuchi:sb,Salam:sp,Lam:qa}.

As usual we split
$\tilde{\Gamma}_{\mathit{cl}}[\Phi,\mathcal{C},\bar{\mathcal{C}}]
= \tilde{\Gamma}_{\mathit{bil}}[\Phi,\mathcal{C},\bar{\mathcal{C}}]
+\tilde{\Gamma}_{\mathit{int}}[\Phi,\mathcal{C},\bar{\mathcal{C}}]$
into the bilinear part 
\begin{align}
\tilde{\Gamma}_{\mathit{bil}}[\Phi,\mathcal{C},\bar{\mathcal{C}}]
= -\frac{1}{2} \Phi_i (\Delta^{-1})^{ij} \Phi_j 
+ \frac{\hbar}{\mathrm{i}} \bar{\mathcal{C}}^i 
\mathcal{C}_i
\end{align}
and an interaction part
$\tilde{\Gamma}_{\mathit{int}}[\Phi,\mathcal{C},\bar{\mathcal{C}}]$,
in which the fields are replaced by functional derivatives with
respect to the sources. Then the functional integration can (formally)
be performed and yields
\begin{align}
Z[J] = \mathcal{N}' \exp\Big( \frac{\mathrm{i}}{\hbar} 
J^i \Omega_i\Big[\frac{\hbar}{\mathrm{i}}
\frac{\partial}{\partial J}\Big] \Big)\;
&
\exp\Big( \frac{\mathrm{i}}{\hbar} 
\tilde{\Gamma}_{\mathit{int}}\Big[\frac{\hbar}{\mathrm{i}}
\frac{\partial}{\partial J}, \frac{\hbar}{\mathrm{i}}
\frac{\partial}{\partial \bar{\mathcal{J}}}, 
\pm \frac{\hbar}{\mathrm{i}}
\frac{\partial}{\partial \mathcal{J}}\Big] \Big) \;
\nonumber
\\
& \hspace*{3em} \times 
\exp\Big( \frac{\mathrm{i}}{2\hbar} J^i \Delta_{ij} J^j 
-  \Big(\frac{\mathrm{i}}{\hbar}\Big)^2 \bar{\mathcal{J}}^i
\mathcal{J}_i \Big)\bigg|_{\mathcal{J}=\bar{\mathcal{J}}=0}\;.
\label{Z1}
\end{align}
The source $J^i$ in front of $\Omega_i$ is external and therefore must
not be differentiated. We can write $J^i$ however as
$(\Delta^{-1})^{ij}\Phi_j$ with $\Phi_j = \frac{\hbar}{\mathrm{i}}
\frac{\delta}{\delta J^j}$ and correct the error due to 
contractions of $J^j$ with other sources. One type of these
contractions is given by a loop of these $J^i\Omega_i$ vertices
in the form 
\begin{align}
\frac{\delta \Omega_{i_1}}{\delta \Phi_{i_2}}   
\frac{\delta \Omega_{i_2}}{\delta \Phi_{i_3}} \dots 
\frac{\delta \Omega_{i_{n-1}}}{\delta \Phi_{i_n}}
\frac{\delta \Omega_{i_n}}{\delta \Phi_{i_1}}\;.      
\label{Cghost}
\end{align}
These loops cancel exactly the ghost loops, because the ghost vertices
are given by $\bar{\mathcal{C}}^i \frac{\delta \Omega_i}{\delta
  \Phi_j} \mathcal{C}_j$ and the ghost propagator equals $1$. Next a
single $J^i \Omega_i$ vertex can be contracted with
$\tilde{\Gamma}_{\mathit{int}}$ to give $\Omega_i 
\frac{\tilde{\Gamma}_{\mathit{int}}}{\delta \Phi_i}$. This new vertex
can further be contracted, as well as the
$\Omega_i(\Delta^{-1})^{ij}\Phi_j$ vertex, and we finally get
\begin{align}
Z[J] = \mathcal{N}' \exp \frac{\mathrm{i}}{\hbar} \Big(&
\Gamma_{\mathit{bil}}[\Phi-\Omega[\Phi-\Omega[\Phi-\dots]]]
- \Gamma_{\mathit{bil}}[\Phi]
\nonumber
\\*
& + \tilde{\Gamma}_{\mathit{int}}[\Phi-\Omega[\Phi-\Omega[\Phi-\dots]]]\Big)
\bigg|_{\Phi\mapsto \frac{\hbar}{\mathrm{i}} \frac{\delta}{\delta J}} 
\exp\Big( \frac{\mathrm{i}}{2\hbar} J^i \Delta_{ij} J^j \Big)\;.
\label{Z11}
\end{align}
Recalling $\Phi'=\Phi+\Omega[\Phi]$ and
$\Gamma[\Phi+\Omega[\Phi]]=\tilde{\Gamma}[\Phi]$, (\ref{Z11})
simplifies to the formula obtained by a direct computation of
(\ref{path}), i.e.\ without the change of variables (\ref{redef}),
\begin{align}
Z[J] &= \mathcal{N}' 
\exp\Big( \frac{\mathrm{i}}{\hbar} 
\Gamma_{\mathit{int}}\Big[\frac{\hbar}{\mathrm{i}}
\frac{\partial}{\partial J}\Big] \Big) 
\exp\Big( \frac{\mathrm{i}}{2\hbar} J^i
\Delta_{ij} J^j \Big)\;.
\label{Z2}
\end{align}
The equivalence of (\ref{Z1}) and (\ref{Z2}) was of course expected.
We are, however, interested in a different question.  It is clear that
(\ref{Z2}) yields the (general) Green's functions of [Case II], but
how can we relate it to the Green's functions of
\cite{Wulkenhaar:2001sq} and [Case I]? 

To answer this question we pass to the generating functional 
\begin{align}
Z_c[J]=\frac{\hbar}{\mathrm{i}} \ln Z[J]
\end{align}
of connected Green's functions and by Legendre transformation to the
generating functional 
\begin{align}
\Gamma[\Phi_{\mathit{cl}}] = Z_c[J] - J^i \Phi_{i,\mathit{cl}}
\label{1PI}
\end{align}
of 1PI Green's functions, where $J^i$ has to be replaced by the
inverse solution of
\begin{align}
\Phi_{i,\mathit{cl}} = \frac{\delta Z_c[J]}{\delta J^i}\;.
\label{1PIcl}
\end{align}
In this way $\Gamma[\Phi_{\mathit{cl}}]$ is obtained as a formal sum
over $\ell$-loop Feynman graphs. The model studied in
\cite{Wulkenhaar:2001sq} is given by the \emph{subset} of Feynman
graphs corresponding to (\ref{Z1}) but \emph{without closed
  ($\mathcal{C},\bar{\mathcal{C}}$)-ghost loops and without the
  vertices involving $\Omega$}. The [Case I] Feynman graphs are
obtained by leaving out the fermionic part of the $\Omega$-vertex and
the corresponding ghosts. We show now that 1PI Graphs in $Z[J]$
involving a single $\Omega$-vertex result in a field redefinition, but
this property does not extend to higher order in $\Omega$.

The 1PI-part of $Z_c[J]$ which is at most linear in $\Omega$ has the
form 
\begin{align}
Z^{\mathrm{1PI},\mathit{lin}(\Omega)}_c[J] = \frac{1}{2} J^i \Delta_{ij} J^j 
+ \tilde{\Gamma}_{\mathit{int}}[\Delta{\cdot} J] 
+ \tilde{\Gamma}_{\mathit{eff}}^{(\ell \geq 1)} [\Delta{\cdot} J] 
+ J^i \Omega_{\mathit{eff},i}^{(\ell\geq 0)}[\Delta{\cdot} J] \;,
\label{Z1PI}
\end{align}
where $(\Delta{\cdot}J)_i = \Delta_{ij} J^j$. All ($\ell{\geq} 1$)-loop
1PI graphs without the $\Omega$-vertex are contained in
$\Gamma_{\mathit{eff}}^{(\ell \geq 1)}$ and all 1PI-graphs involving
the $\Omega$-vertex are contained in
$\Omega_{\mathit{eff},i}^{(\ell\geq 0)}$. All graphs are built with
the $\tilde{\Gamma}_{\mathit{int}}$ vertices and $(\mathcal{\bar{C}},
\mathcal{C})$-ghost loops are omitted, assuming the ghost tadpole
$\frac{\delta \Omega_i}{\delta \Phi_i}$ in (\ref{Cghost}) to be zero.
Now we obtain
\begin{align}
\Phi_{i,\mathit{cl}} &= (\Delta{\cdot} J)_i 
+  \Delta_{ij} 
\frac{\delta \tilde{\Gamma}_{\mathit{int}}}{\delta \Phi_j}[\Delta{\cdot} J]
+  \Delta_{ij} 
\frac{\delta \tilde{\Gamma}_{\mathit{eff}}^{(\ell \geq 1)}}{
\delta \Phi_j}[\Delta{\cdot} J]
\nonumber
\\
& +  \Delta_{ij} J^k
\frac{\delta \Omega_{\mathit{eff},k}^{(\ell \geq 0)}}{
\delta \Phi_j}[\Delta{\cdot} J]
+ \Omega_{\mathit{eff},i}^{(\ell \geq 0)} [\Delta{\cdot} J]\;,
\\
(\Delta{\cdot} J)_i^{\mathit{lin}(\Omega)} 
&= \Phi_{i,\mathit{cl}} 
- \Delta_{ij} \frac{\delta \tilde{\Gamma}_{\mathit{int}}}{
\delta \Phi_j} [\Phi_{\mathit{cl}}]
- \Delta_{ij} 
\frac{\delta \tilde{\Gamma}_{\mathit{eff}}^{(\ell \geq 1)}}{
\delta \Phi_j} [\Phi_{\mathit{cl}}]
\nonumber
\\*[-1ex]
&
+ \Delta_{ij} 
\frac{\delta^2 \tilde{\Gamma}_{\mathit{int}}}{\delta \Phi_j \delta \Phi_k} 
[\Phi_{\mathit{cl}}] \,
\Omega_{\mathit{eff},k}^{(\ell \geq 0)}[\Phi_{\mathit{cl}}]
+ \Delta_{ij} 
\frac{\delta^2 \tilde{\Gamma}_{\mathit{eff}}^{(\ell \geq 1)}}{
\delta \Phi_j \delta \Phi_k} [\Phi_{\mathit{cl}}]\,
\Omega_{\mathit{eff},k}^{(\ell \geq 0)}[\Phi_{\mathit{cl}}]
\nonumber
\\*[-1ex]
&
- \Omega_{\mathit{eff},i}^{(\ell \geq 0)} [\Phi_{\mathit{cl}}]
- \Delta_{ij} J^k
\frac{\delta \Omega_{\mathit{eff},k}^{(\ell \geq 0)}}{\delta \Phi_j}
[\Phi_{\mathit{cl}}]
+\text{1PR-terms}\;,
\label{DJ}
\\
\Gamma^{\mathit{lin}(\Omega)} [\Phi_{\mathit{cl}}] &= \Big(
\tilde{\Gamma}_{\mathit{eff}}^{(\ell \geq 0)}[\Delta{\cdot}J] 
- (\Delta{\cdot} J)_i \frac{\delta \tilde{\Gamma}_{\mathit{int}}}{
\delta \Phi_i}[\Delta{\cdot}J]
- (\Delta{\cdot} J)_i \frac{\delta
  \tilde{\Gamma}_{\mathit{eff}}^{(\ell \geq 1)}}{\delta
  \Phi_i}[\Delta{\cdot}J]
\nonumber
\\*
& 
\hspace*{15em} - (\Delta {\cdot} J)_i J^k 
\frac{\delta \Omega_{\mathit{eff},k}^{(\ell \geq 0)}}{
\delta \Phi_i}[\Delta{\cdot}J]
\Big)^{\mathrm{1PI},\mathit{lin}(\Omega)}
\nonumber
\\*
&= \tilde{\Gamma}_{\mathit{eff}}^{(\ell \geq 0)}[\Phi_{\mathit{cl}}] 
-\frac{\delta \tilde{\Gamma}_{\mathit{eff}}^{(\ell \geq 0)}}{
\delta \Phi_i}[\Phi_{\mathit{cl}}] \, 
\Omega_{\mathit{eff},i}^{(\ell \geq 0)}[\Phi_{\mathit{cl}}]\;.
\label{Geff}
\end{align}
Terms like $\big(\Phi_{i,\mathit{cl}} 
-\Omega_{\mathit{eff},i}^{(\ell \geq 0)}\big)\frac{\delta
  \Gamma_{\mathrm{int}}}{\delta \Phi_i}[\Phi_{\mathit{cl}}]$ cancel
via the direct occurrence in the first line of (\ref{Geff}) and the
substitution of (\ref{DJ}) in $\Gamma_{\mathit{bil}}[\Delta{\cdot}J]=
-\frac{1}{2} (\Delta{\cdot}J)\Delta^{-1} (\Delta{\cdot}J)$. The final
result (\ref{Geff}) shows that graphs involving the $\Omega$-vertices
in (\ref{Z1}) linearly are a field redefinition. In our 
$(n{=}1,\ell{=}1)$ loop calculation the $\Omega$-vertices contribute
already with $\ell=1$, therefore, the effect at total loop order 1 is
expected to be
\begin{align}
\Omega_{\mathit{eff},i}^{(\ell =1)} \frac{\delta
  \Gamma_{\mathit{cl}}}{\delta \Phi_i}\;,
\end{align}
which is exactly the difference of the loop calculations of 
\cite{Wulkenhaar:2001sq}, [Case I] and [Case II]. 

Taking graphs with more than one $\Omega$-vertex into account, the
difference of the cases cannot be a field redefinition any longer.
Namely, there is now a graph $J^{i_1}\cdots J^{i_n} \Omega^{(\ell \geq
  1)}_{\mathit{eff},i_1\dots i_n}[\Delta{\cdot}J]$ in the
generalisation of (\ref{Z1PI}), which gives the term $(1{-}n)
J^{i_1}\cdots J^{i_n} \Omega^{(\ell \geq 1)}_{\mathit{eff},i_1\dots
  i_n}[\Delta{\cdot}J]$ in $\Gamma$. The free sources $J^{i_k}$ are
now replaced e.g.\ by $\frac{\delta\Gamma_{\mathit{eff}}}{\delta
  \Phi_k}$ and thus lead to 1PI graphs where the $\Omega$-vertices
become \emph{inner}. These graphs cannot be reached by field
redefinitions, which are \emph{outer}. In conclusion, we expect at
order $\theta^2$ that the differences between
\cite{Wulkenhaar:2001sq}, [Case I] and [Case II] are no longer field
redefinitions.

In principle there are also the
($\mathcal{C},\bar{\mathcal{C}}$)-ghost loops to take into account.
However, the corresponding ghost propagator equals 1 and the ghost
couplings are polynomial in momenta and masses. If there are no
sub-divergences, all ghost loops vanish trivially, at least in analytic
and dimensional regularisation. Accordingly, if the
($\mathcal{C},\bar{\mathcal{C}}$)-ghost vertices are renormalisable,
the ($\mathcal{C},\bar{\mathcal{C}}$)-ghosts give no contribution at
all.

\section{Discussion}
\label{sec6}

In this paper we have continued the quantum analysis of the
Seiberg-Witten map first carried out in \cite{Bichl:2001nf,
  Bichl:2001cq, Wulkenhaar:2001sq}. We have analysed $\theta$-expanded
non-commutative QED, which happens to be the easiest non-commutative
model to study in this context. In contrast to
\cite{Wulkenhaar:2001sq}, where both bosonic and fermionic sectors
were $\theta$-expanded via the Seiberg-Witten differential equations,
we have analysed in this paper the two cases where 
\begin{itemize}
\item[(I)] 
only the bosonic sector is expanded via the Seiberg-Witten map and 

\item[(II)] neither the bosonic nor the fermionic sectors 
are expanded via the Seiberg-Witten map.
\end{itemize}
We have found that up to field redefinitions the outcome of all three
approaches is identical. We can summarise our picture about the
Seiberg-Witten map as follows:
\begin{itemize}
\item The Seiberg-Witten expansion must be seen as a true (physical)
  expansion of the fields in a gauge theory, which is performed {\it
    prior} to quantisation. Otherwise (expanding after the
  quantisation) ghosts and $\Omega$-vertices generated due to the
  change of integration variables would contribute to the loop
  calculation and lead to the same result as without the
  Seiberg-Witten map.
  
\item At first order in $\theta$ no difference between
  $\theta$-expanded quantum field theories with and without
  Seiberg-Witten map is expected (apart from problems with the choice
  of the gauge group, which we ignore here). Our one-loop QED
  calculations confirm this.
  
\item $\theta$-expanded gauge theory can {\it not} be expected to be
  stable under quantisation because divergences will appear already at
  first order in $\theta$, for the reason that no symmetry is known
  which rules them out. At first order in $\theta$ the additional
  terms added to the initial action in order to have enough freedom to
  absorb these divergences are the same when using the Seiberg-Witten
  map or leaving it out.

\item At second order in $\theta$ there will be substantial differences
  between $\theta$-expansion with or without Seiberg-Witten map due to
  contributions of the $\Omega$-vertices (and possibly
  non-renormalisable ghost sub-divergences).
\end{itemize}

The most important result of this paper is perhaps that {\it iff} one
insists on analysing $\q$-expanded (Abelian) gauge theories involving
fermions one {\it must} add the term
\begin{align}
\nonumber
g_e \int \!d^4x\;  
\mathrm{i} \theta^{\alpha\beta}
\hat{\bar{\psi}} \star \gamma^{\mu\nu}_{~~\alpha} 
\hat{D}^{adj}_\beta \hat{F}_{\mu\nu}\star \hat{\psi}
\end{align}
to the non-expanded initial action. Also, the fermion masses should be
introduced via a Higgs mechanism.

Let us finally stress that it is not yet possible to make definite
conclusions towards renormalisability of $\q$-expanded models. It
appears that explicit loop calculations at second order in $\q$ are
needed, these are however not easily accessible due to the enormous
volume of calculations involved.

\section*{Acknowledgement}

We would like to thank Peter Schupp for giving us the initial idea to
investigate the mixed case where bosons are Seiberg-Witten expanded
whereas fermions are not.  We would like to thank Paolo Ascheri,
Branislav Jur\v{c}o, Peter Schupp, Harold Steinacker and Julius Wess
for stimulating discussions and hospitality during our visit at the
Physics Department of the University of Munich.

\addcontentsline{toc}{section}{\numberline{}References}

\end{document}